\documentstyle[epsf,12pt,subfigure]{article}

\begin{document}
\title{The Dynamics of Image Processing by Feature Maps in the Primary Visual Cortex}
\author{
	Ted Hesselroth \and Klaus Schulten\\
	Department of Physics and  Beckman Institute\\
	University of Illinois\\
	Urbana, Illinois 61801\\[1ex]
	}

\maketitle
\eject

{\em Abstract.}
{\small The operational characteristics of a linear neural network image processing system based on the brain's vision system are investigated. The final stage of the network consists of edge detectors of various orienations arranged in a feature map, corresponding to the primary visual cortex, or V1. The  lateral geniculate nucleus is modeled as a preprocessing stage. Excitatory forward and inhibitory backward connections exist between the LGN and V1. By a method of  reconstructing the input images in terms of V1 activities, the simulations show that images can be faithfully represented in V1 by the proposed network. The signal-to-noise ratio of the image is improved by the representation, and compression ratios of well over two-hundred are possible. Lateral interacations between V1 neurons sharpen their orientational tuning. We further study the dynamics of the processing, showing that the rate of decrease of the error of the representation is maximized for the receptive fields used, and we develop a Fokker-Planck equation for a more detailed prediction of the error value vs. time. Finally, we show how the eigenvalues of the covariance matrix of the inputs can be employed to predict the rate of error decrease.}

\section{Introduction}
\setcounter{table}{0}

The companion paper to this one may be found at http://arxiv.org/abs/q-bio.NC/0505011.

In that work 
 we described a neural network which developed oriented edge detectors and a feature map very closely resembling those found in the primate visual system. We postulated that inhibitory backward connections from V1 to the LGN play an important role in the transfer of information from LGN to V1, and showed that the rate of information transfer from the LGN to V1 is maximized with the receptive fields developed.

Here we use the mature network from that research to investigate the dynamics of image processing in the visual system. The present work is divided into two parts.

The first set of analyses focuses on the properties of the representation obtained by the network. We will show that natural images can be faithfully represented by the collection of edge detectors developed under hebbian learning rules, that the signal to noise ratio of the information is improved by the feedback mechanism described, and that a sharpening of orientation tuning of V1 neurons results from the lateral interactions between V1 neurons. On the way, we explore the compression possibilities of the algorithms used in the simulations. 

The second half of the investigation is a study of the actual dynamics of the image processing done by the network. It is demonstrated that the signals in both the LGN and V1 are transient as expected, and time series of the neural activities are compared to experimental results.  We also show how that network behavior can be predicted from a knowledge of the eigenvalues of the covariance matrix of the inputs.

\subsection{Non-orthogonal Representations}



If the receptive fields of cortical neurons can detect all or most of the essential features of an image, then it should be possible to reconstruct the image from the responses of the feature detectors. Such a reconstruction is useful for investigating various properties of the encoding of an image into V1 activities. A natural choice for the reconstruction is to use the receptive fields of the V1 neurons as a set of basis vectors within the space of  possible images. 

Each feature detecting neuron in V1 has only a limited range of image elements that it can respond to. Therefore, each feature detector can encode only a very small amount of information about the visual scene. The simplicity of the receptive fields is compensated for in the brain by the use of large numbers of neurons. The processing system must be able to combine these many  simple elements into a coherent, detailed interpretation of the image. Because the receptive fields are of a simple form and each has a relatively large extent in the visual field, there must be a considerable overlap between them to cover the image features that must be represented. Taken as a set of basis elements for the representation of the image, the receptive fields of V1 neurons thus form a non-orthogonal basis.

If the set of basis elements is orthonormal, the coefficients of the representation are simply the inner products of the image with the basis vectors. When the basis elements are non-orthogonal,  there are various ways of calculating the coefficients, typically towards the end of minimizing the squared error between the representation and  the input image. Numerous works have been published concerning the Gabor scheme of image representation, which also utilizes a set of non-orthogonal coefficients, and these works are relevant to the present investigation. In the Gabor scheme, the basis elements are trigonometric functions along particular directions in space, masked by gaussian functions. Some authors have proposed dual functions to find Gabor coefficients by direct integration\cite{BAST81A,PORA88A}. Others employ neural networks which arrive at the coefficients of  the representation of an image. \cite{DAUG88A,PATT92A,GAAL93A}. Some work\cite{LIZH93,LENZ92A,SAKI82A,KAMI93} has also been done on image representation using feature detectors similar to the ones described here, but no previous work, to our knowledge, has utilized feature detectors arranged in a feature map such as is found in the visual cortex.

\section{The Computational Method}

The network is exactly the same as that of the previous chapter, only with mature weight values, and the updating of weight values is not included. For reference purposes, the dynamical equations are described again in the following.

\subsection{Algorithms}

The algorithms for the dynamics of the network areas follows. The variables $x_i$, $y_j$, and $a_k$ are the input a photoreceptor at
position $i$, and activities of a retinal  ganglion cell at position $j$, and a V1 neuron at position $k$, in their  respective neural layers.

Input images are convolved by a difference-of-gaussians function representing the receptive fields of the 
center-surround retinal ganglion cells. We  put
\begin{equation}\label{eq:lgnact}
y_j = \sum_i  g_{ij} \: x_i
\end{equation} 
where $g_{ij}$ is the difference-of-gaussians function.  The $y_j$ are the initial activities in the LGN layer. There are feedforward connections from the LGN to V1, and lateral connections between V1 neurons. The activity of
neurons in layer V1 in terms of the activities in the LGN layer is given by:
\begin{equation}\label{eq:v1actnolat2}
a_k = \sum_{j,k} w_{jk} \: y_j
\end{equation}
where $w_{jk}$ is the set of connection strengths, .e.g., between neuron $j$ in the LGN and neuron $k$ in V1. The lateral interactions between V1 neurons may also be included, in which case we have
\begin{equation}\label{eq:v1act2}
a_l = \sum_{j,k} h_{kl} \: w_{jk} \: y_j
\end{equation}
with $h_{kl}$ for the lateral connection between V1 neurons $k$ and $l$. That is, the lateral interaction is computed after the forward interaction, i.e., the forward and lateral interactions are applied in  succession.

The backward connections are inhibitory and decrease the activity of the LGN relay neuron:
\begin{equation}\label{eq:inhib}
\dot{y}_j = -\sum_k  w_{jk} \: a_k
\end{equation}
The backward connections have the same values as the forward connections, since they learn from the same inputs. 

An image is convolved with the difference-of-gaussians functions to form an input LGN image, which is stored in an array. A V1 neuron is chosen and its activities are calculated from \ref{eq:v1actnolat2}. In some simulations, the lateral interaction may then be applied to the neuron's neighbors. Then the effect of the neuron's activity on the LGN through the backward connections is applied to the LGN activity. This defines one step. The process is repeated until each of the V1 neurons has been chosen exactly once. This defines one epoch. 

\subsection{The Error Measure}

In order to analyse the representation of the image in the V1 layer, the image is reconstructed from the V1 activities. Integration of equation \ref{eq:inhib} yields a formula for the LGN activities $y_j(t)$ in terms of the initial activities $y_j(0)$ and the V1 activities.
\begin{equation}\label{eq:yval}
y_j(t) = y_j(0) - \sum_k  w_{jk} \: \int_{t'=0}^{t}  a_k(t') dt'
\end{equation}
Now, if one defines a error measure in terms of the LGN activities,
\begin{eqnarray}\label{eq:eee}
 E  &=&  \sum_{j }\:  y_j^2\\
\nonumber
\end{eqnarray}
upon inserting the learning rule into the time derivative of $E$,
\begin{eqnarray}\label{eq:dotee}
\dot E  &=&  \sum_{j }\:  y_j \dot{y}_j\\
\nonumber\\
&=&  -\sum_{j} \:  y_j  \: a_l \: w_{jl}\\
\nonumber\\
&=&  -(a_l)^2\\
\nonumber
\end{eqnarray}
it is seen, then, that the error decreases on every step, and thus goes to zero. Therefore the LGN activities must all go to zero, and by eq. \ref{eq:v1actnolat2}, so must the V1 activities. Since the left-hand side of eq. \ref{eq:yval} goes to zero, the dynamical equations indicate that the time integral of the V1 activities are the coefficients of the representation of the LGN image with the connection strength to the V1 neurons as the basis set.

This can be shown another way, by the use of the psuedoinverse.  Define {\bf y} as a vector with components $y_j$ running over the index $j$, similarly {\bf a} as $a_k$ over the components $k$, and  ${\bf w}_k$ as $w_{jk}$ over the index $j$. The matrix ${\bf W}$ be $w_{jk}$  is formed by taking  the vectors ${\bf w}_k$  as its rows.
Then, for example,  equation \ref{eq:v1actnolat2} is written
\begin{equation}\label{eq:v1actnolat3}
{\bf a} = {\bf W} \: {\bf y}
\end{equation}
That is , $a_k$ is the $k^{th}$ component of the product of ${\bf W}$ with the vector {\bf y}. Or, $a_k$ is the inner product of ${\bf w}_k$ with {\bf y}.
With ${\bf W}^+$, as the psuedoinverse of ${\bf W}$, 
\begin{equation}\label{eq:pseudodef}
{\bf W}^+ = ({\bf W}^T {\bf W})^{-1} \: {\bf W}^T
\end{equation}
then, using a matrix identity,
\begin{equation}\label{eq:defwf2}
{\bf W^T \: W}^+ = \alpha \sum_{n=0}^\infty \: {\bf W}^T \: {\bf W} \: (I - \alpha \: {\bf W}^T \: {\bf W})^n
\end{equation}
is the projection onto the subspace spanned by the vectors ${\bf w}_k$. But this is exactly the reconstruction described above, since $(I - \alpha \: {\bf W}^T \: {\bf W})\: {\bf y}$ is the activity of the LGN layer after one iteration, an interation being composed of determining the V1 activities by equation \ref{eq:v1actnolat2} and inhibiting the LGN through the backward inhibitory connections. Multiplying thr result by ${\bf W}$ gives the V1 activities after one epoch, and then by ${\bf W}^T$ gives the reconstruction after one epoch. Taking ${\bf W}^T$ out of the sum, the sum of the first $N$ terms is the sum of the V1 activities from each of the first $N$ epochs, e.g., the "integral" of the V1 activities. Thus the reconstruction method used herein is simply the projection of the input image onto the subspace spanned by the weight vectors. The above also shows that the network presented cannot be improved on for finding the correct coefficients for minimizing the mean squared error, since the pseudoinverse does just that.

\section{Characteristics of the Image Representations Produced by the Network}

The visual system must contend with a number of constraints in its task of representing images. Each V1 neuron has only a limited range of features that it can respond to, and so any representation must be built up from a large number of very simple descriptions. Metabolic energy of the organism is limited, requiring that the total activity of neurons be minimized. Speed of computation is essential in relating to a changing environment. Finally, internal noise generated by the image processing system should not interfere with the image representation.

Let us examine, then, the various physical limitations of the particular processing elements making up the vision system, and determine whether and how the proposed model addresses the requirements they introduce.

\subsection{The Lena Image}

The prime requirement is that the network represent images faithfully. We apply the network to a portrait in order to see if the reconstruction resembles the input image. The input image is considered to be the initial activities of the LGN neurons, the convolution of the original input image. 

\begin{figure}[p] 
\begin{center} 
\begin{tabular}[t]{c} 
\subfigure[]{\parbox{0.45\textwidth}{ 
\epsfxsize=0.45\textwidth 
\epsfysize=0.45\textwidth 
\epsffile{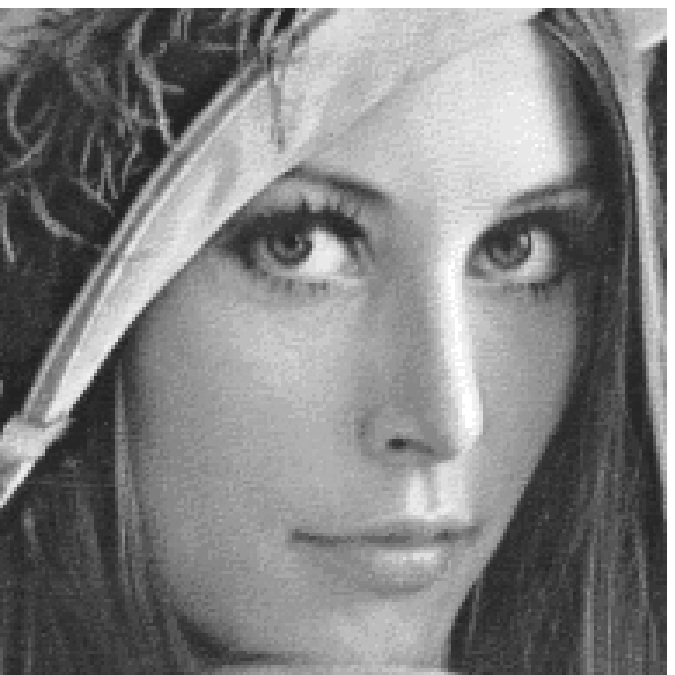} 
} 
} 
\subfigure[]{\parbox{0.45\textwidth}{ 
\epsfxsize=0.45\textwidth 
\epsfysize=0.45\textwidth 
\epsffile{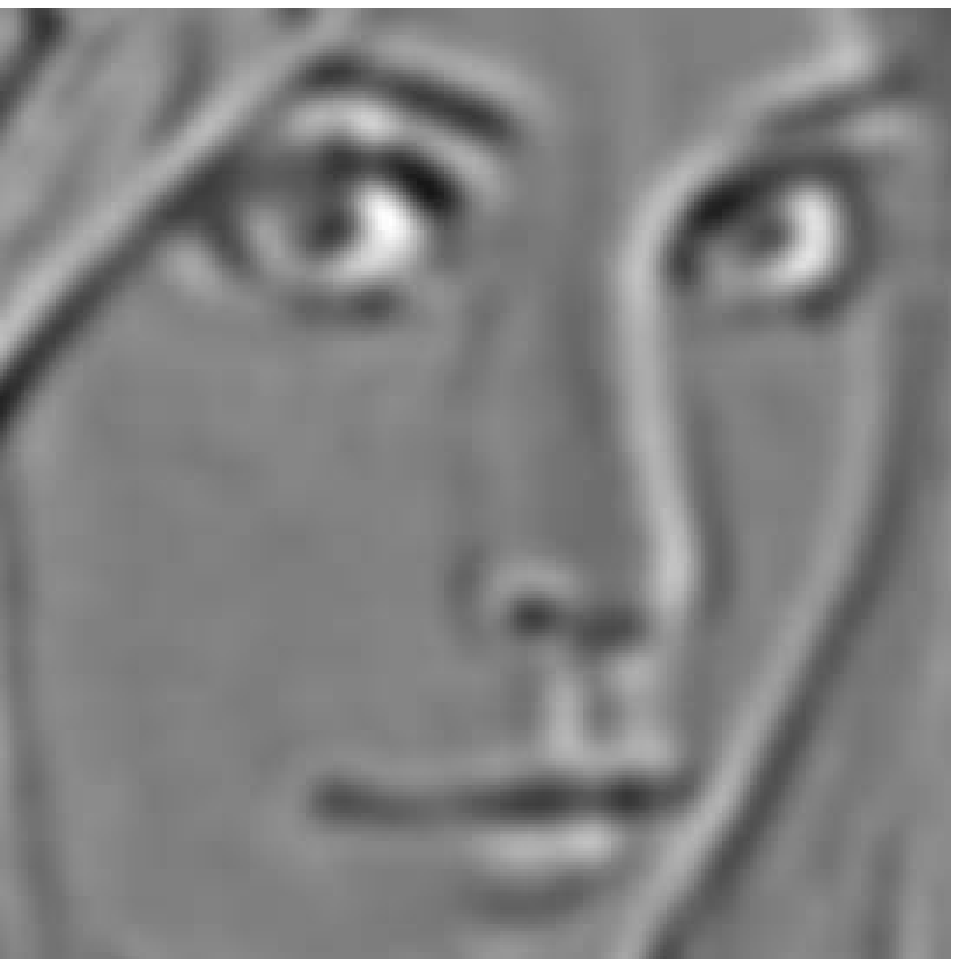}  
} 
} 
\vspace{0.0cm}\\  
\subfigure[]{\parbox{0.45\textwidth}{ 
\epsfxsize=0.45\textwidth 
\epsfysize=0.45\textwidth 
\epsffile{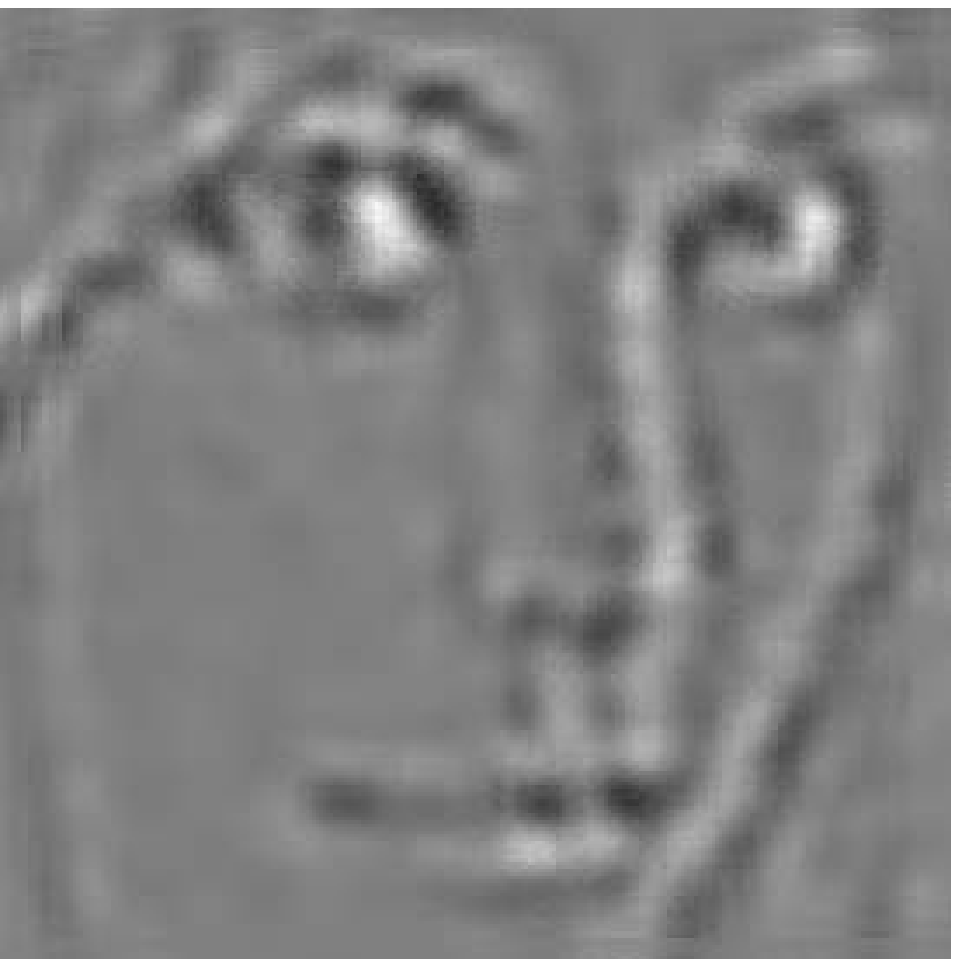} 
} 
} 
\subfigure[]{\parbox{0.45\textwidth}{ 
\epsfxsize=0.45\textwidth 
\epsfysize=0.45\textwidth 
\epsffile{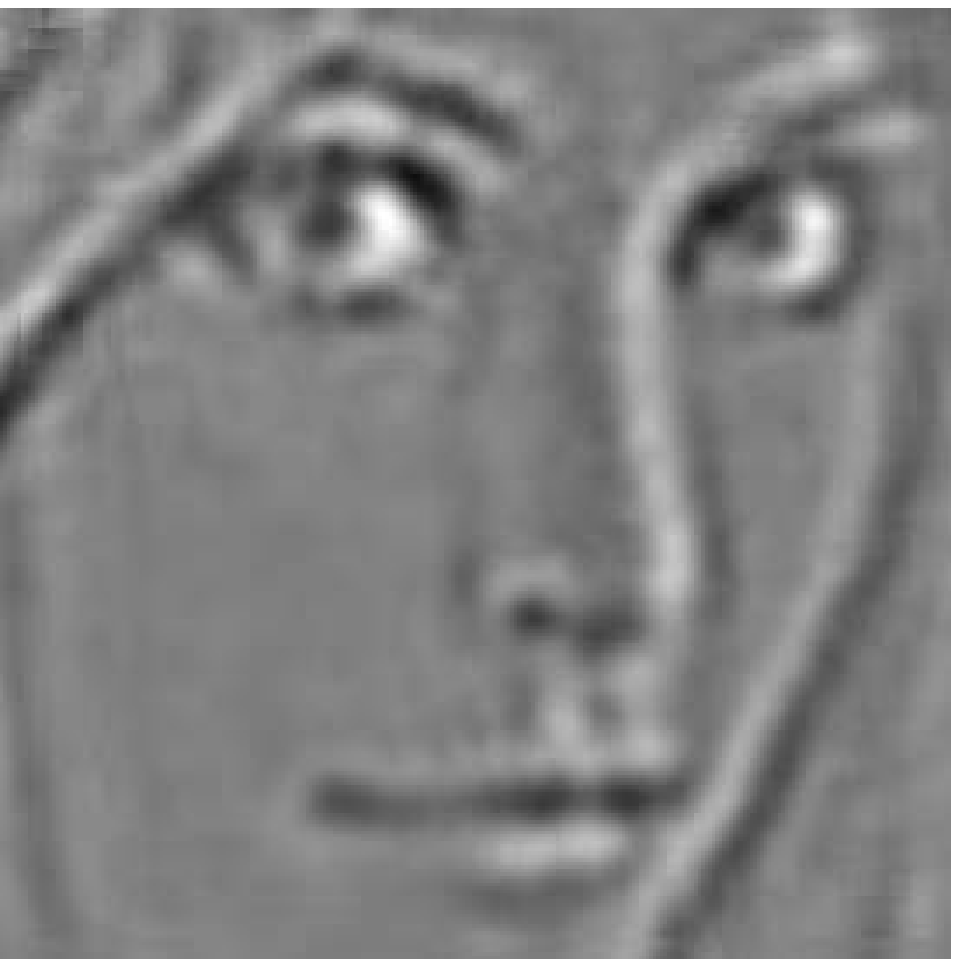} 
} 
} 
\end{tabular} 
\end{center}
\vspace{-0.5cm} 
\caption[]{ a) Original lena image. b) Original lena image to LGN after 
filtering through retina. c) Reconstructed image after 1 epoch d) Reconstructed 
image after 20 epochs.} 
\label{fig:p:lena1} 
\end{figure}

The "lena" image is shown in figure \ref{fig:p:lena1} a. This is the image considered to be input on the retina. The image is convolved through the difference-of-gaussians function which imitates the activity of the retinal ganglion cells. The result forms the input to the network and is shown in figure \ref{fig:p:lena1} b.

After one epoch of processing, the reconstructed image begins to resemble the input image, fig. \ref{fig:p:lena1} c. After twenty epochs, the reconstruction approximates the original, fig. \ref{fig:p:lena1} d. This shows that the information from the input image has been is well-represented by the integrals of the V1 activities.

\subsection{The Effect of Lateral Interactions}

\begin{figure}[p] 
\begin{center} 
\begin{tabular}[t]{c} 
\subfigure[]{\parbox{0.45\textwidth}{ 
\epsfxsize=0.45\textwidth 
\epsfysize=0.45\textwidth 
\epsffile{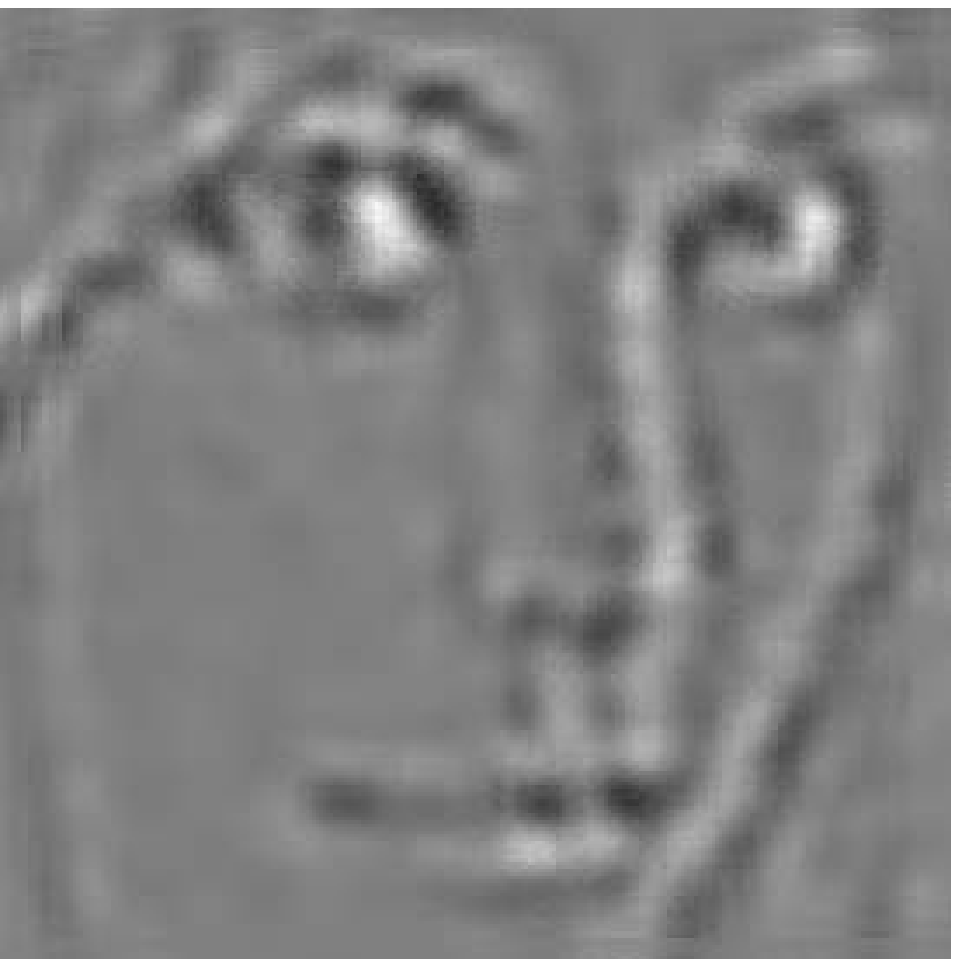}  
} 
} 
\subfigure[]{\parbox{0.45\textwidth}{ 
\epsfxsize=0.45\textwidth 
\epsfysize=0.45\textwidth 
\epsffile{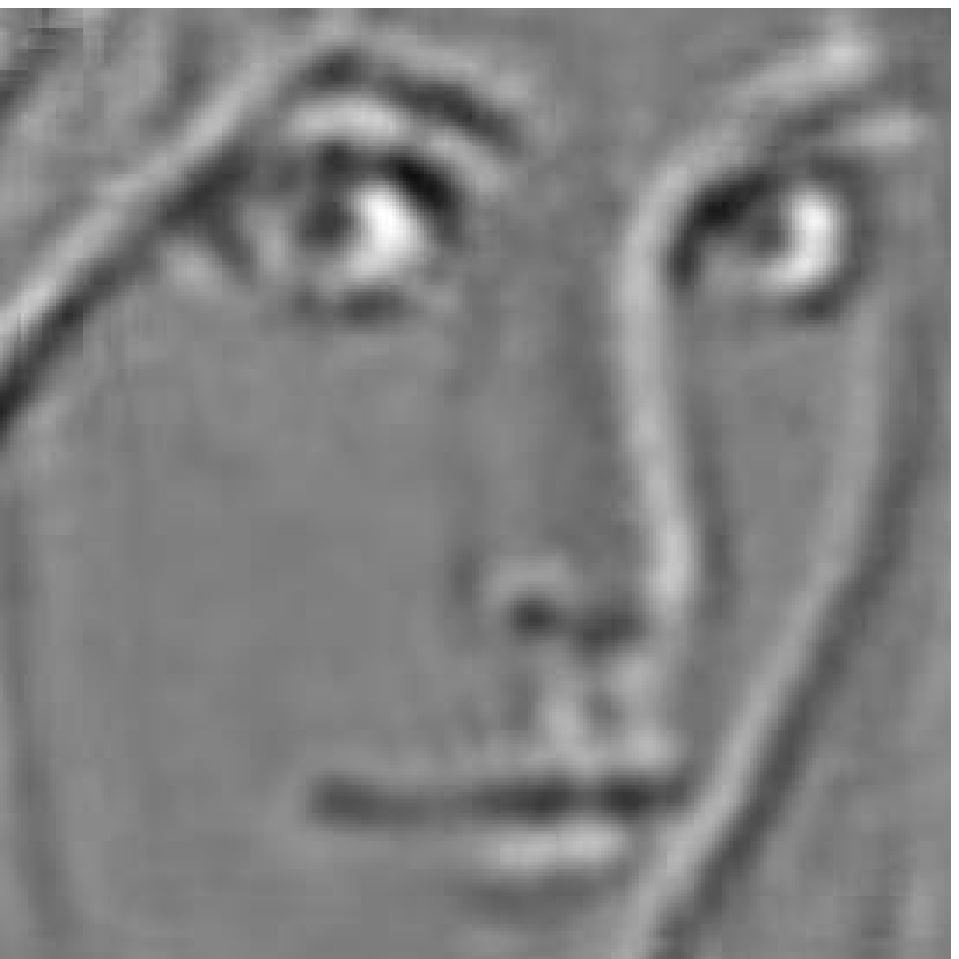} 
} 
} 
\vspace{0.0cm}\\  
\subfigure[]{\parbox{0.45\textwidth}{ 
\epsfxsize=0.45\textwidth 
\epsfysize=0.45\textwidth 
\epsffile{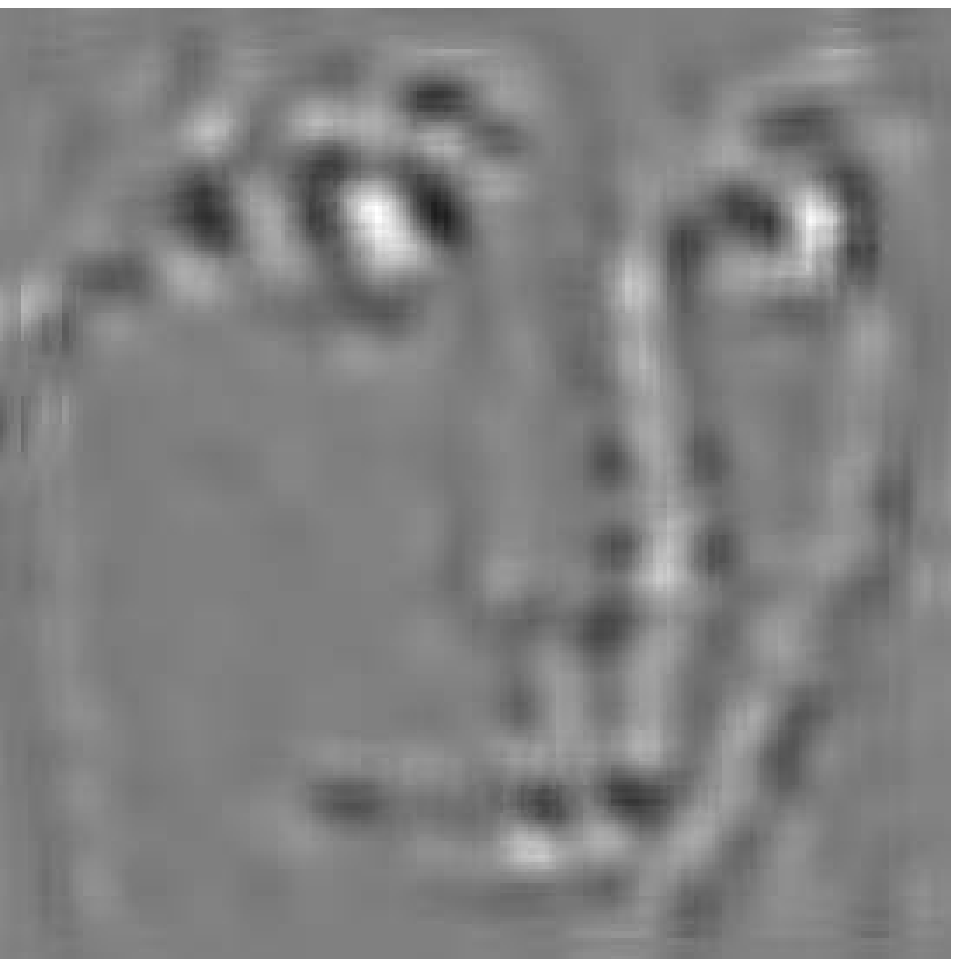}  
} 
} 
\subfigure[]{\parbox{0.45\textwidth}{ 
\epsfxsize=0.45\textwidth 
\epsfysize=0.45\textwidth 
\epsffile{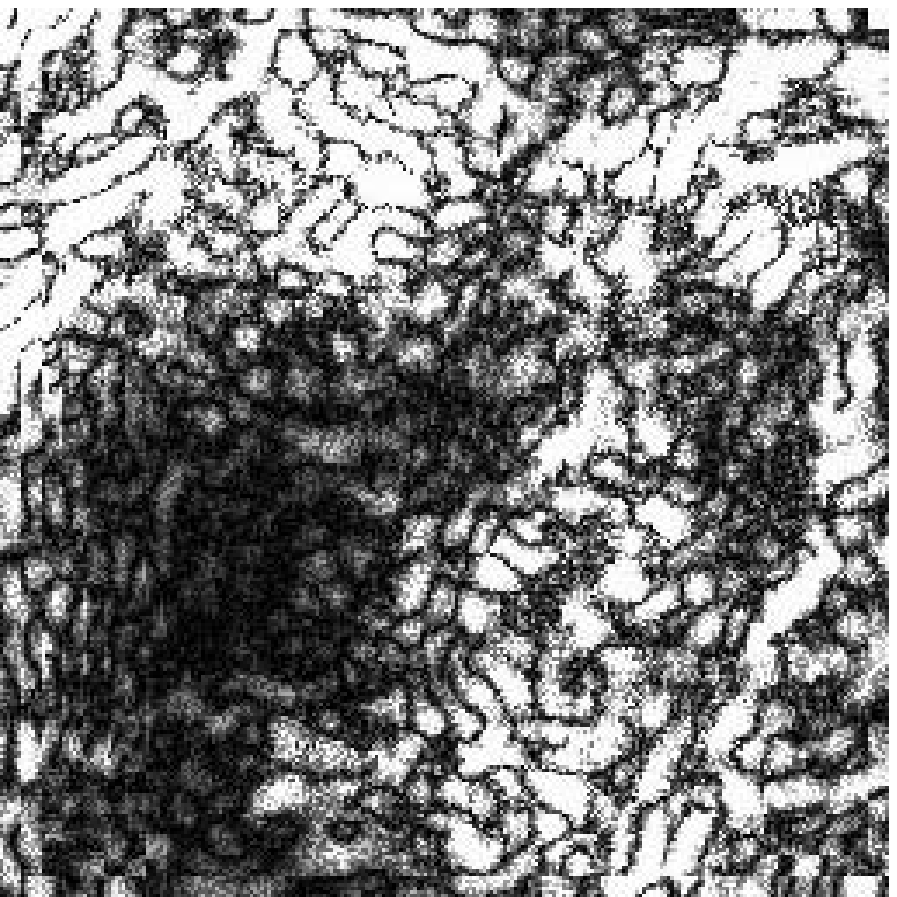} 
} 
} 
\end{tabular} 
\end{center}
\vspace{-0.5cm} 
\caption[]{ a) Reconstructed image after 1 epoch with lateral interaction. b) 
Reconstructed image after 20 epochs with lateral interaction. c) Reconstructed 
image after 1 epochs with no inhibitory feedback. d) Absolute values of 
activities in V1 after twenty steps, constrast enhanced, with inhibitory 
feedback and no lateral interactions.} 
\label{fig:p:lena2} 
\end{figure}

When lateral interactions between V1 neurons are included, the results after one and after twenty epochs are shown in figs. \ref{fig:p:lena2} a and \ref{fig:p:lena2} b respectively. We find that the lateral interactions do improve the image quality slightly. The reader is reminded that the lateral interactions were included in the original simulations of the development of the feature map in order to produce the spatial features of the map itself, not for the sake of image processing. There may be additional functions of the lateral interactions in the realm of hyperacuity which are not apparent in the image presented in this example. Later, we test the contribution of the lateral interactions on the phenomenon of sharpening of orientation tuning with a simple edge image.

\subsection{Reconstruction Without Inhibitory Feedback}

When the processing is done without inhibitory feedback (and no lateral interactions), only one epoch need be calculated, since the signal is feedforward only. The reconstructed image is shown in figure \ref{fig:p:lena2} c. The image is not only inferior to the image after twenty epochs with feedback, it is slightly worse than the processing with feedback after one epoch. This shows that with the reconstruction method that was used, the feedback scheme proposed substantially improves the representation of the image.

\subsection{The Activities in V1}

The V1 activities after twenty epochs, with inhibitory feedback and without lateral interactions, are shown in figure \ref{fig:p:lena2} d. The absolute values are shown since the sign of any weight vector in the network is arbitrary. This image is difficult to interpret directly, hence the use of the reconstruction technique to analyse the representation. Though the brain probably does not reconstruct the image as we have done, the reconstruction does show that the image information is indeed in the V1 activities and could be accessed by higher stages of processing in the brain. 

The notable feature of the image of V1 activities if the presence of continuous lines of low activity. It is understandable that they are there: at a singular point there are edge-detecting weight vectors of all orientations; therefore at least one of the neurons will have zero activity. Since away from the singular points the feature map is continuous and the image is also continuous, one would expect continuity in the lines of constant activity value in the V1 activity image. Hence the lines connect singular points together, or they may connect other lines of zero activity. 

In the LGN, such lines correspond to edges in images. We suppose that when the neurons supplying information about edges are quiet, other kinds of neurons, such as those supplying information about shading or color, can win any neural competition that may exist. This corresponds to known psychophysical data which implies that such information indeed is taken from the border area between image regions. Perhaps there are neurons in V1 which supply information, such as on textures, in a similar fashion.

\subsection{Walsh Patterns}\label{walshsec}

Walsh patterns are images consisting of rectangular tilings of black and white areas. The spatial frequency of the tiling may be varied in both the horizontal and vertical directions. Walsh patterns were originally used as a set of basis elements for image representation. They are used here for the sake of comparison with neuroscience experiments. As simple figures with well-defined edges, they are also appropriate for evaluating the performance of the image processing network.

\begin{figure}[p] 
\begin{center} 
\begin{tabular}[t]{c} 
\subfigure[]{\parbox{0.45\textwidth}{ 
\epsfxsize=0.45\textwidth 
\epsfysize=0.45\textwidth 
\epsffile{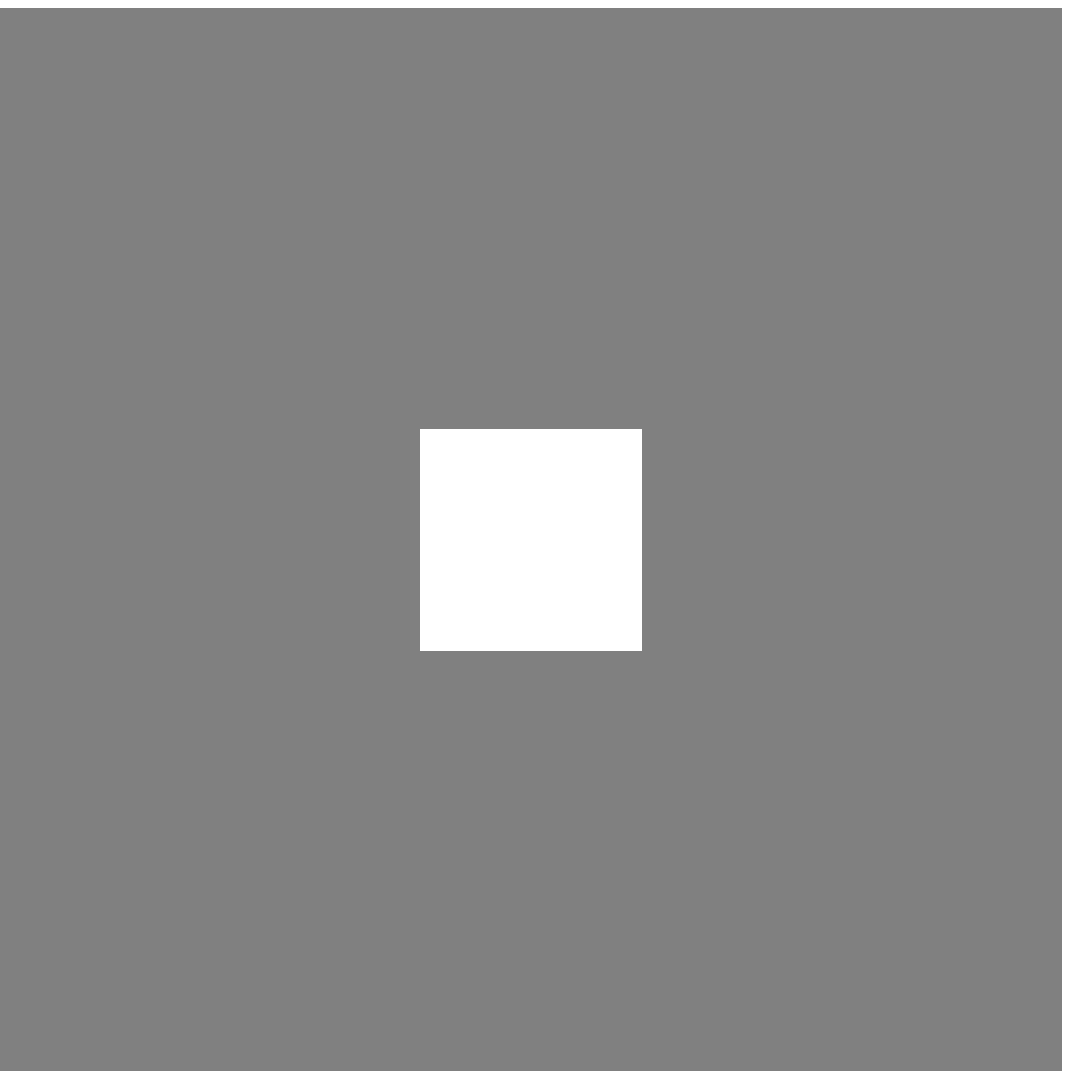}  
} 
} 
\subfigure[]{\parbox{0.45\textwidth}{ 
\epsfxsize=0.45\textwidth 
\epsfysize=0.45\textwidth 
\epsffile{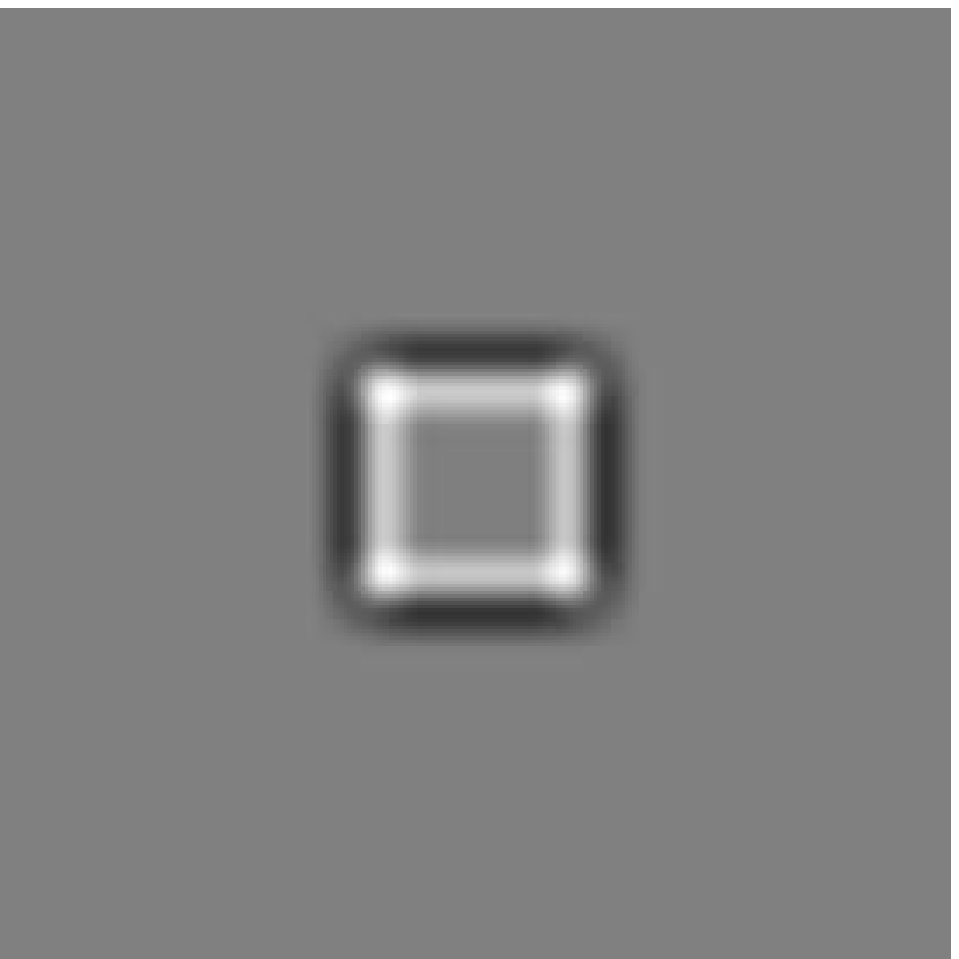}  
} 
} 
\vspace{0.0cm}\\  
\subfigure[]{\parbox{0.45\textwidth}{ 
\epsfxsize=0.45\textwidth 
\epsfysize=0.45\textwidth 
\epsffile{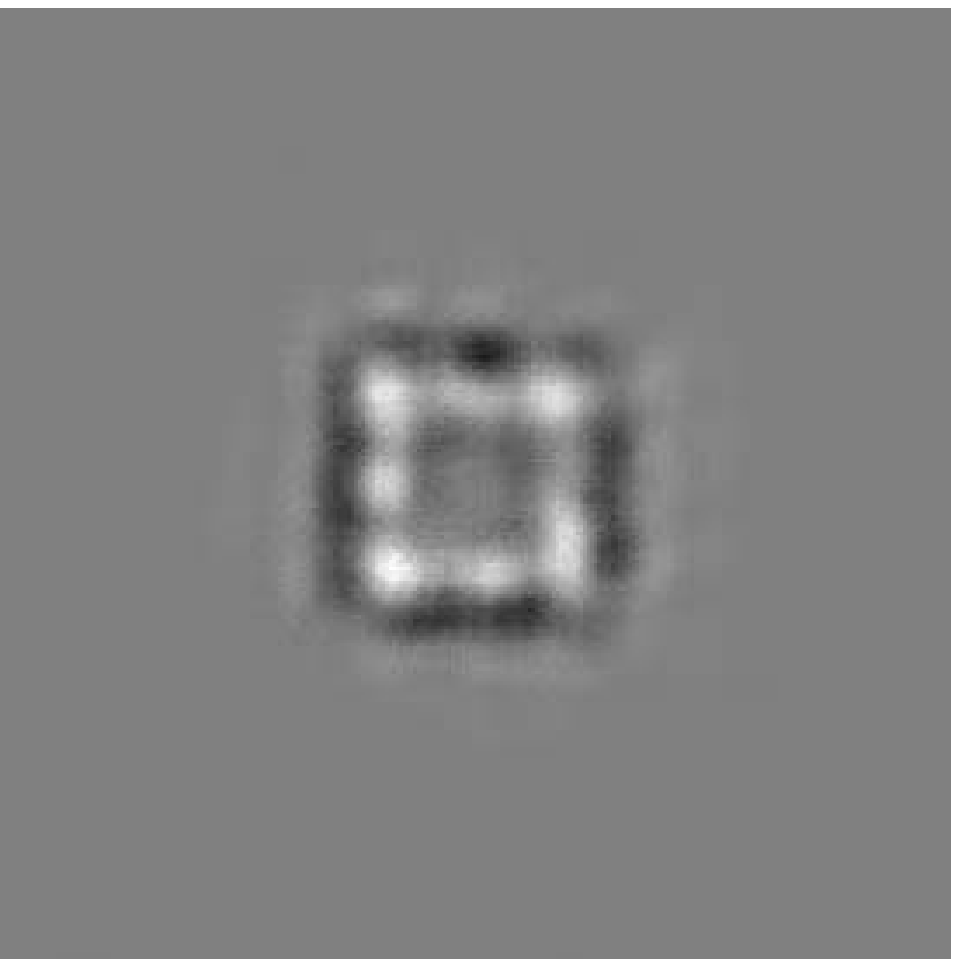} 
} 
} 
\subfigure[]{\parbox{0.45\textwidth}{ 
\epsfxsize=0.45\textwidth 
\epsfysize=0.45\textwidth 
\epsffile{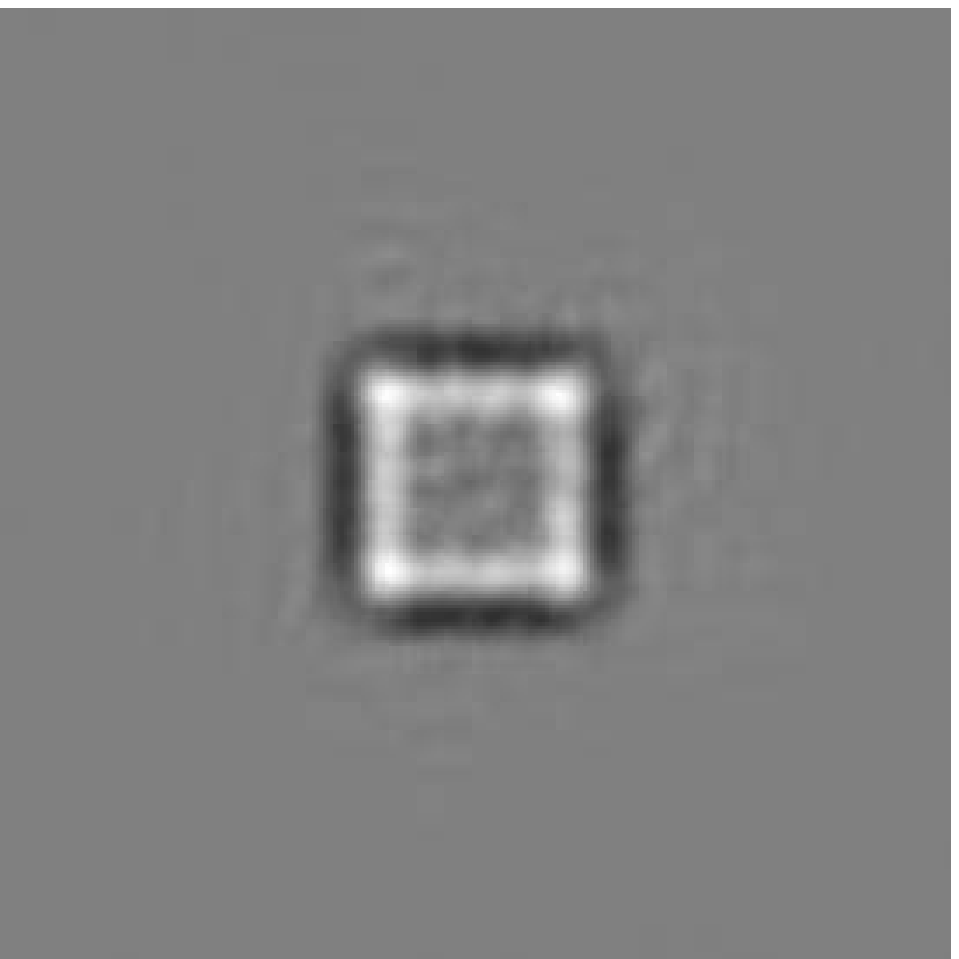}  
} 
} 
\end{tabular} 
\end{center}
\vspace{-0.5cm}
\caption[]{ a) Original walsh pattern (1,1) image. b) Original LGN input for 
walsh image. c) Reconstruction of walsh image after 1 epoch. d) Reconstruction 
of walsh image after 20 epochs.} 
\label{fig:p:walsh11} 
\end{figure}

Walsh pattern (1,1) is a tiling of frequency 1 in both directions. The original image, its convolution through the difference-of-gaussians function, and the reconstructed images at one epoch and twenty epochs are shown in figure \ref{fig:p:walsh11}. In the reconstruction after one epoch, only a few V1 neurons, those almost directly over the edges of the pattern, are contributing to the reconstruction. The size and shape of the V1 receptive fields are visible because of the separation of the neurons involved. At twenty epochs, the representation in V1 is much more distributed, and the neural activities are combining in such a way as to form edges, even where there are no single neurons with edges at those locations. Thus the reconstruction is much more smooth.

\begin{figure}[p] 
\begin{center} 
\begin{tabular}[t]{c} 
\subfigure[]{\parbox{0.45\textwidth}{ 
\epsfxsize=0.45\textwidth 
\epsfysize=0.45\textwidth 
\epsffile{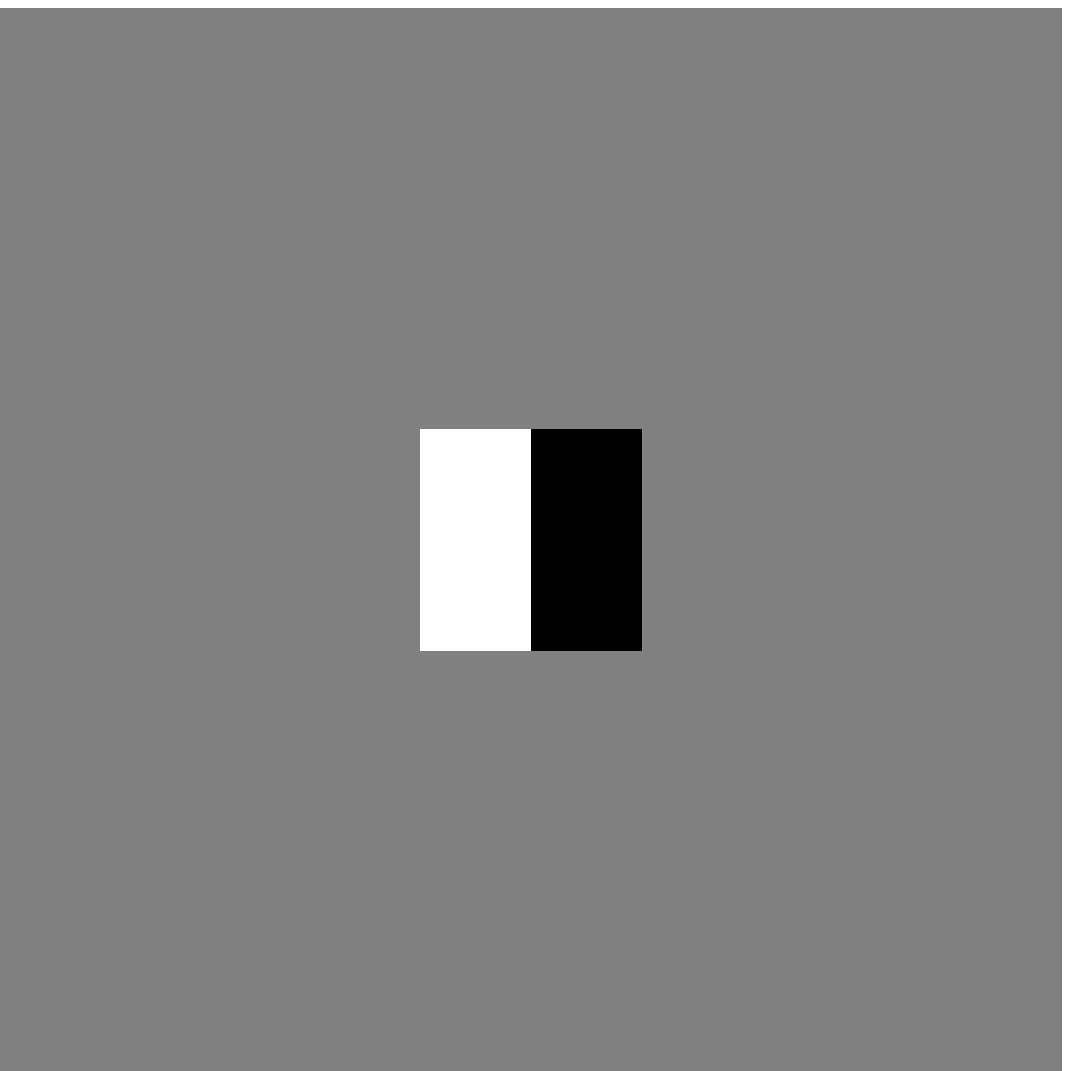}  
} 
} 
\subfigure[]{\parbox{0.45\textwidth}{ 
\epsfxsize=0.45\textwidth 
\epsfysize=0.45\textwidth 
\epsffile{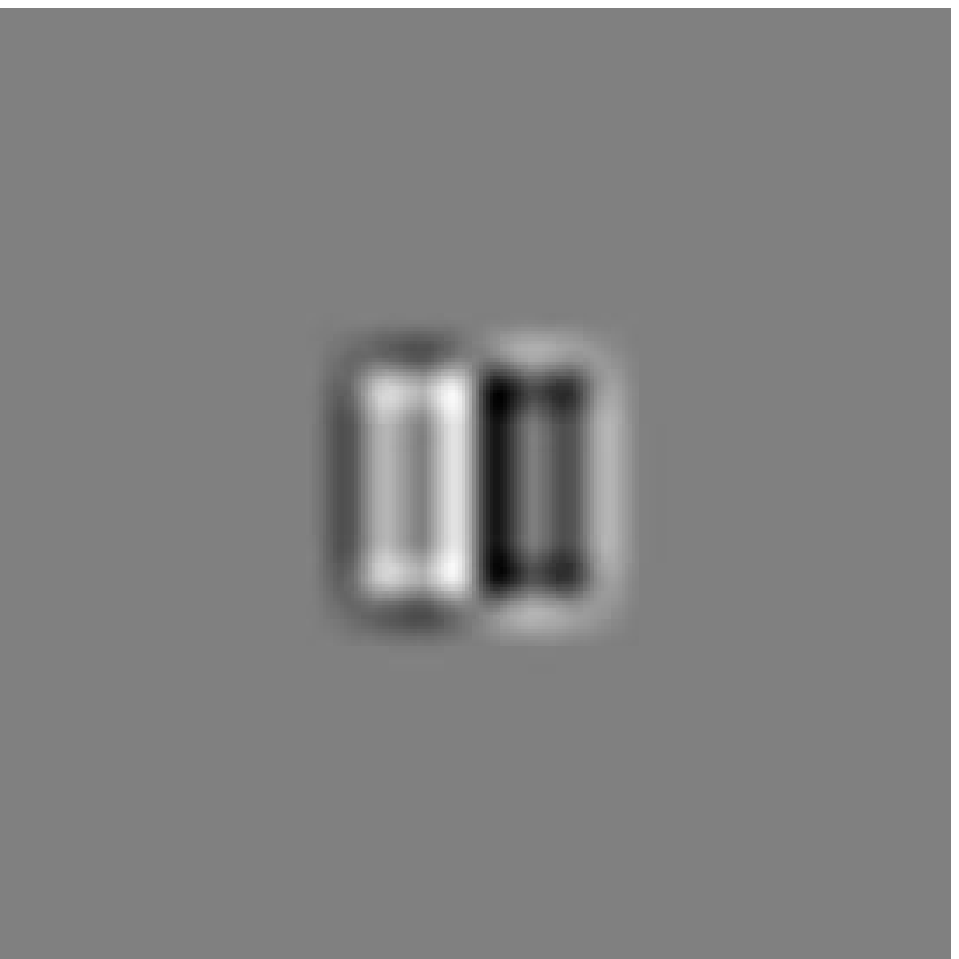}  
} 
} 
\vspace{0.0cm}\\  
\subfigure[]{\parbox{0.45\textwidth}{ 
\epsfxsize=0.45\textwidth 
\epsfysize=0.45\textwidth 
\epsffile{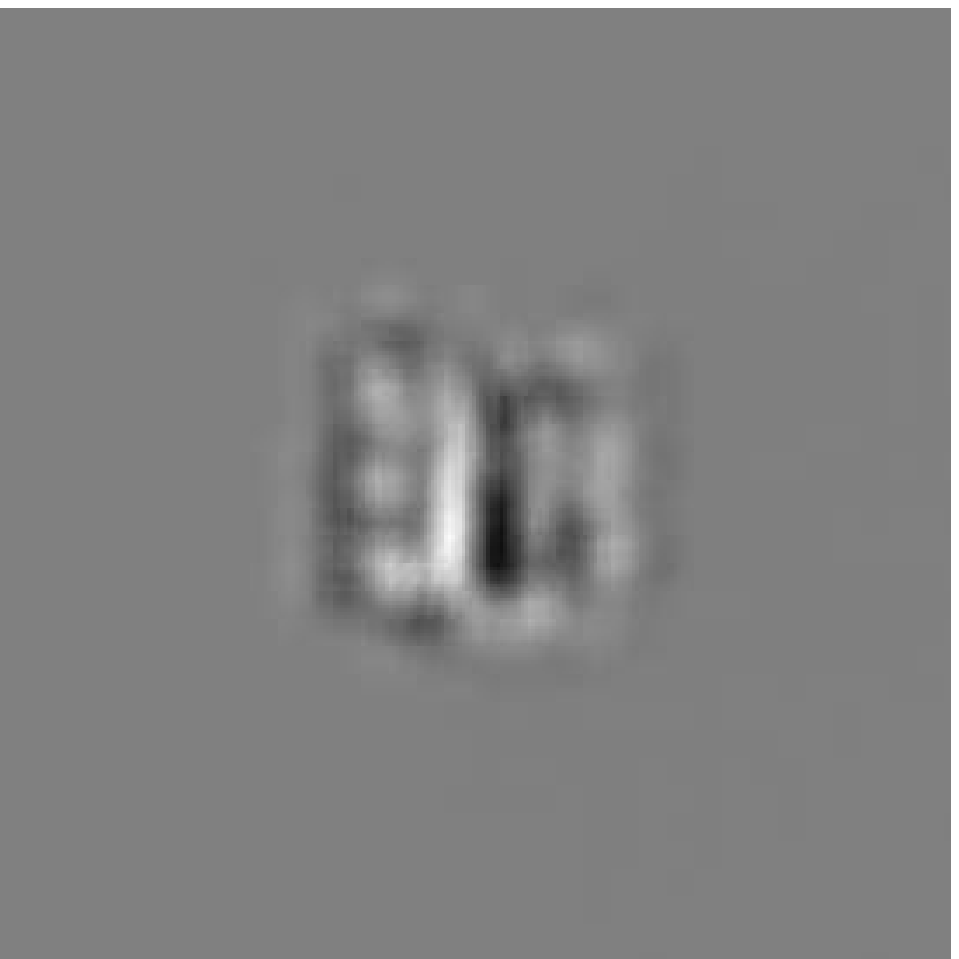}  
} 
} 
\subfigure[]{\parbox{0.45\textwidth}{ 
\epsfxsize=0.45\textwidth 
\epsfysize=0.45\textwidth 
\epsffile{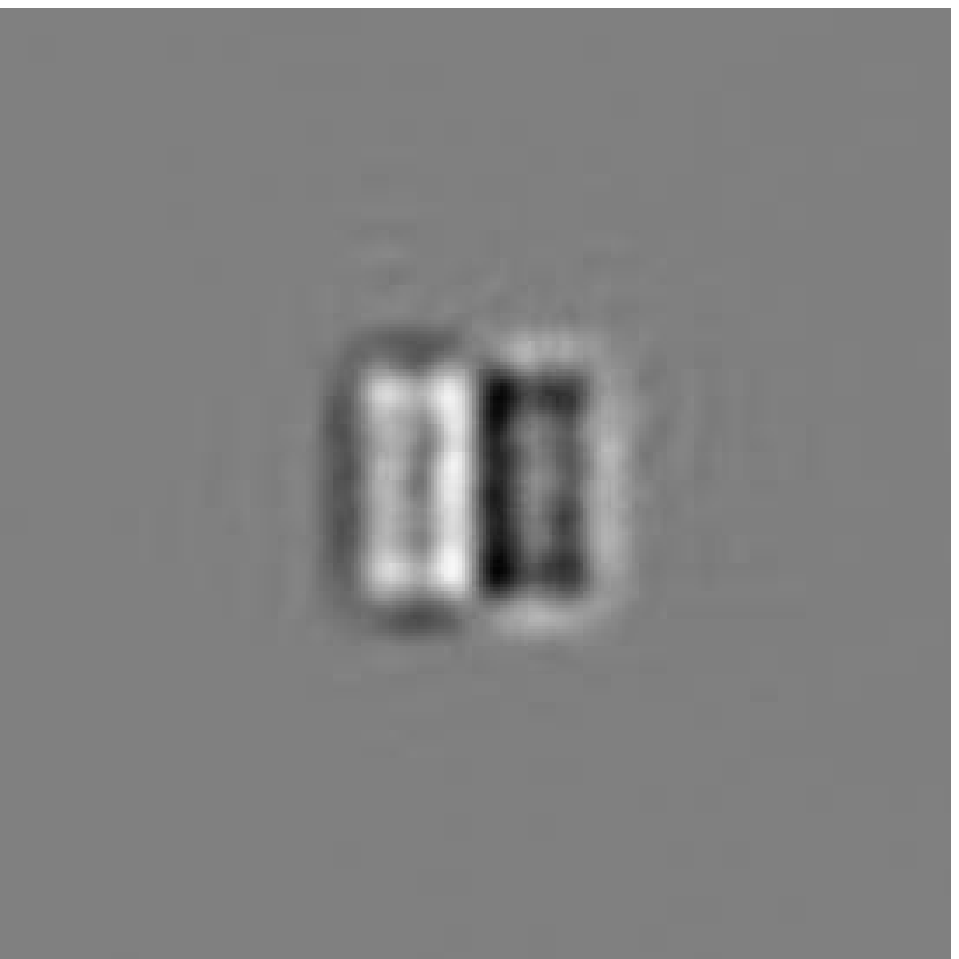}  
} 
} 
\end{tabular} 
\end{center}
\vspace{-0.5cm} 
\caption[]{ a) Original walsh pattern (1,2) image. b) Original LGN input for 
walsh image. c) Reconstruction of walsh image after 1 epoch. d) Reconstruction 
of walsh image after 20 epochs.} 
\label{fig:p:walsh12} 
\end{figure}

\begin{figure}[p] 
\begin{center} 
\begin{tabular}[t]{c} 
\subfigure[]{\parbox{0.45\textwidth}{ 
\epsfxsize=0.45\textwidth 
\epsfysize=0.45\textwidth 
\epsffile{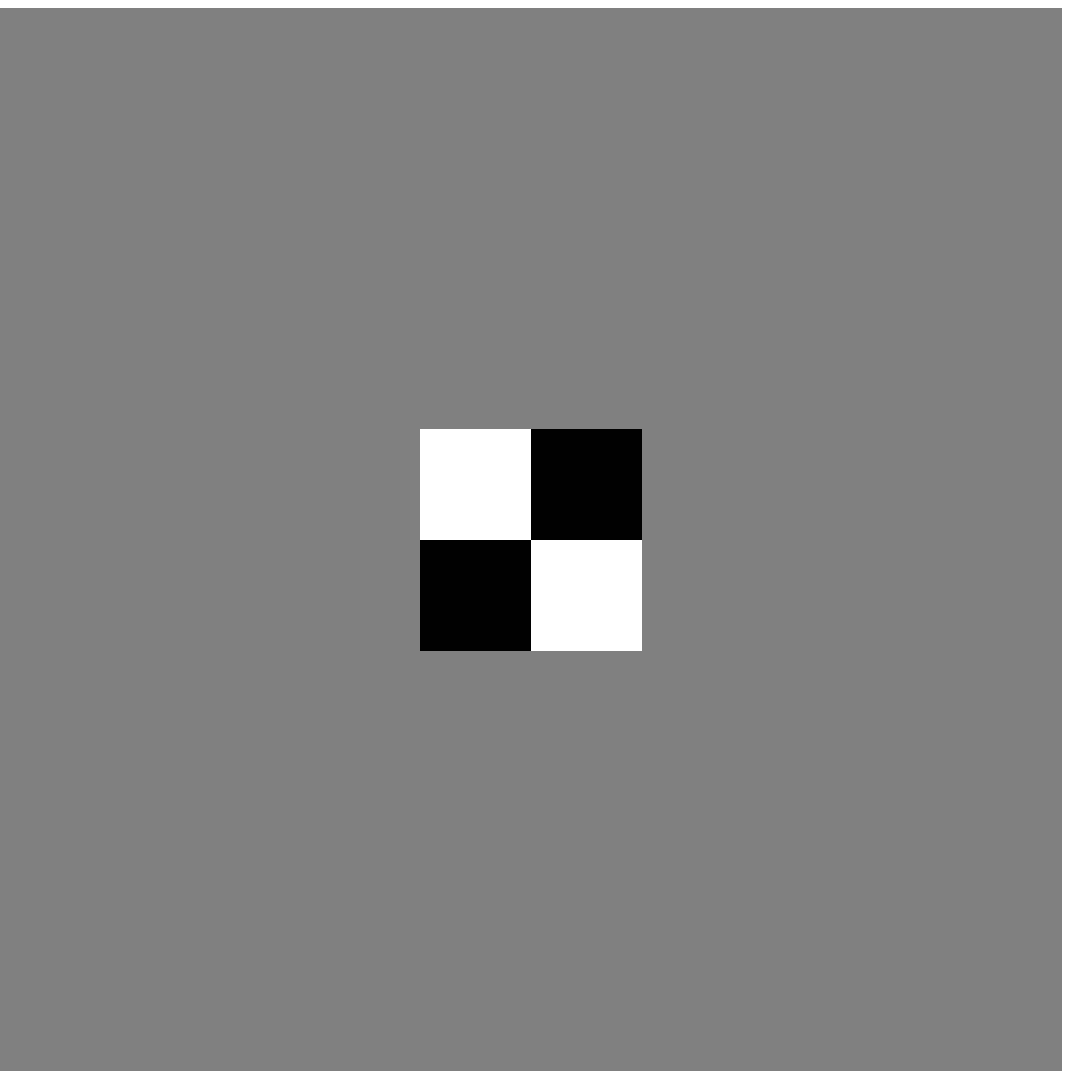} 
} 
} 
\subfigure[]{\parbox{0.45\textwidth}{ 
\epsfxsize=0.45\textwidth 
\epsfysize=0.45\textwidth 
\epsffile{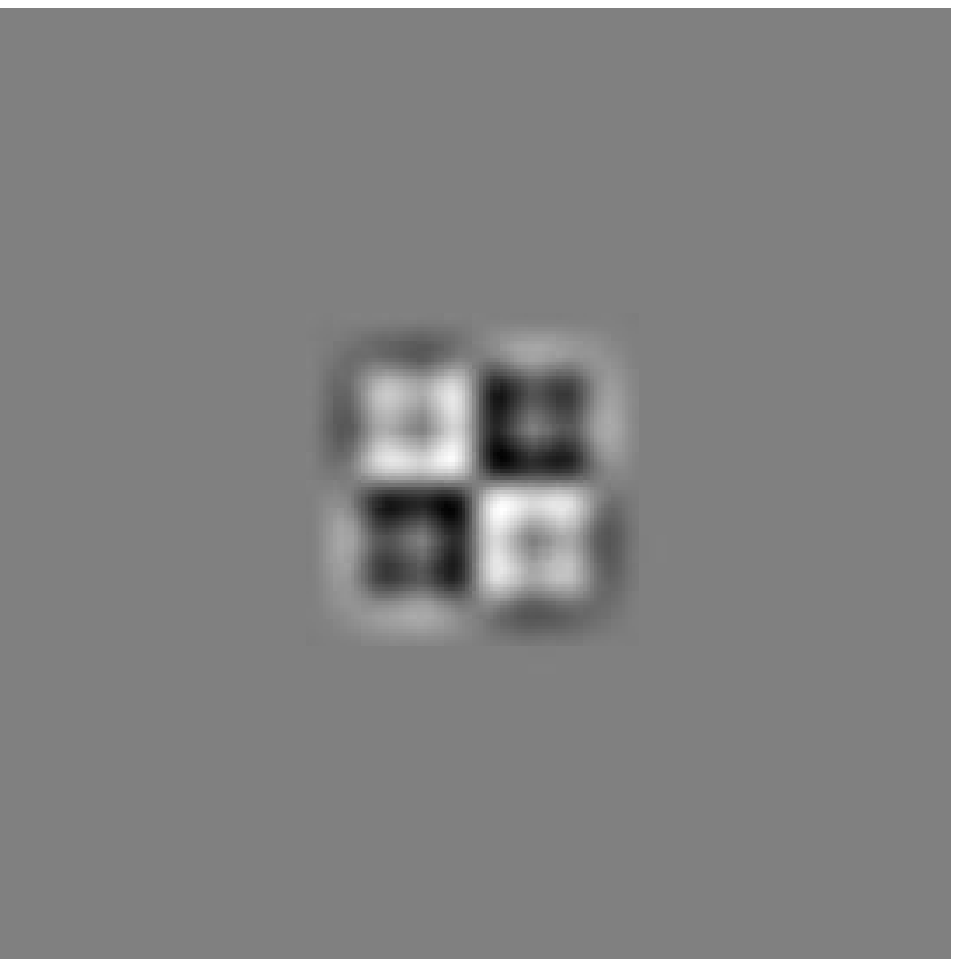}  
} 
} 
\vspace{0.0cm}\\  
\subfigure[]{\parbox{0.45\textwidth}{ 
\epsfxsize=0.45\textwidth 
\epsfysize=0.45\textwidth 
\epsffile{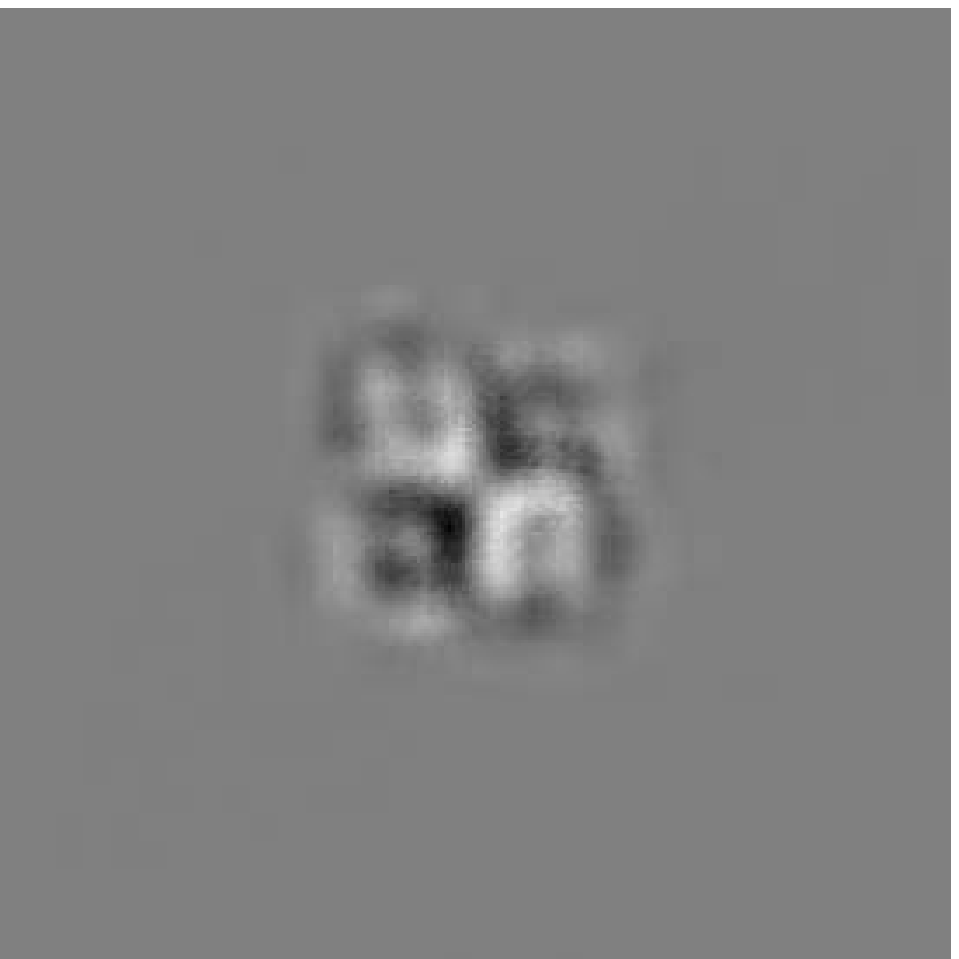} 
} 
} 
\subfigure[]{\parbox{0.45\textwidth}{ 
\epsfxsize=0.45\textwidth 
\epsfysize=0.45\textwidth 
\epsffile{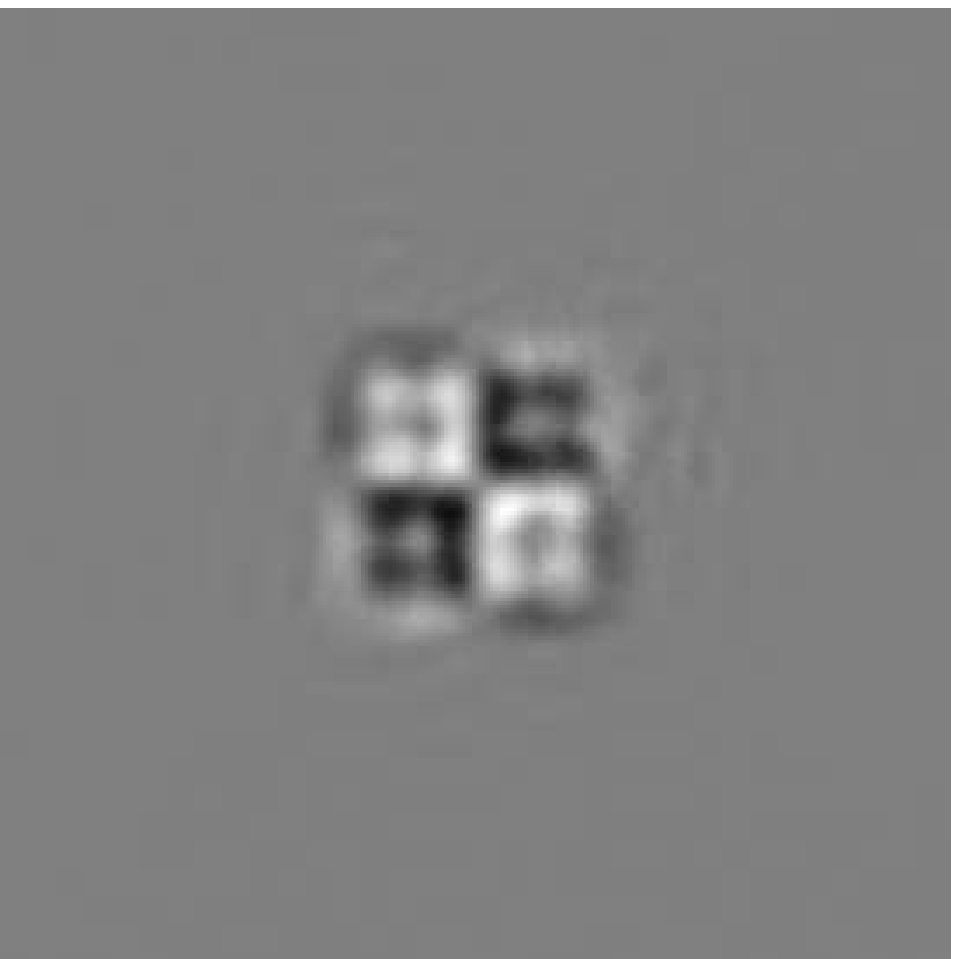}  
} 
} 
\end{tabular} 
\end{center}
\vspace{-0.5cm}
\caption[]{ a) Original walsh pattern (2,2) image. b) Original LGN input for 
walsh image. c) Reconstruction of walsh image after 1 epoch. d) Reconstruction 
of walsh image after 20 epochs.} 
\label{fig:p:walsh22} 
\end{figure}

Walsh patterns (1,2), (2,2), and (4,4) are presented in figures \ref{fig:p:walsh12}, \ref{fig:p:walsh22}, and \ref{fig:p:walsh44}. The center edge in figure \ref{fig:p:walsh12} d is sharper because there is greater contrast in the original image there. Figure \ref{fig:p:walsh44} c shows an interesting effect which is confirmed in experimental results in a later section. The reconstructed image after one epoch is closer to the original compared to the results for simpler images, because the receptive field size and shape happens to be close to the size and shape of the image features to be represented. Therefore the image is well-represented with comparatively less processing. As shown later, this results in a decrease in LGN and V1 activities that is more rapid than expected, and is seen experimentally as well.


\subsection{Improvements in the Signal to Noise Ratio}

Noise is pervasive in the vision processing system. Neurons may fire spontaneously, or there may be voltage fluctuations along afferent fibers, random release of neurotransmitter, and random voltage fluctuations in the receiving neuron. Thus, some immunity to noise must be built into the system. 

\begin{figure}[p] 
\begin{center} 
\begin{tabular}[t]{c} 
\subfigure[]{\parbox{0.45\textwidth}{ 
\epsfxsize=0.45\textwidth 
\epsfysize=0.45\textwidth 
\epsffile{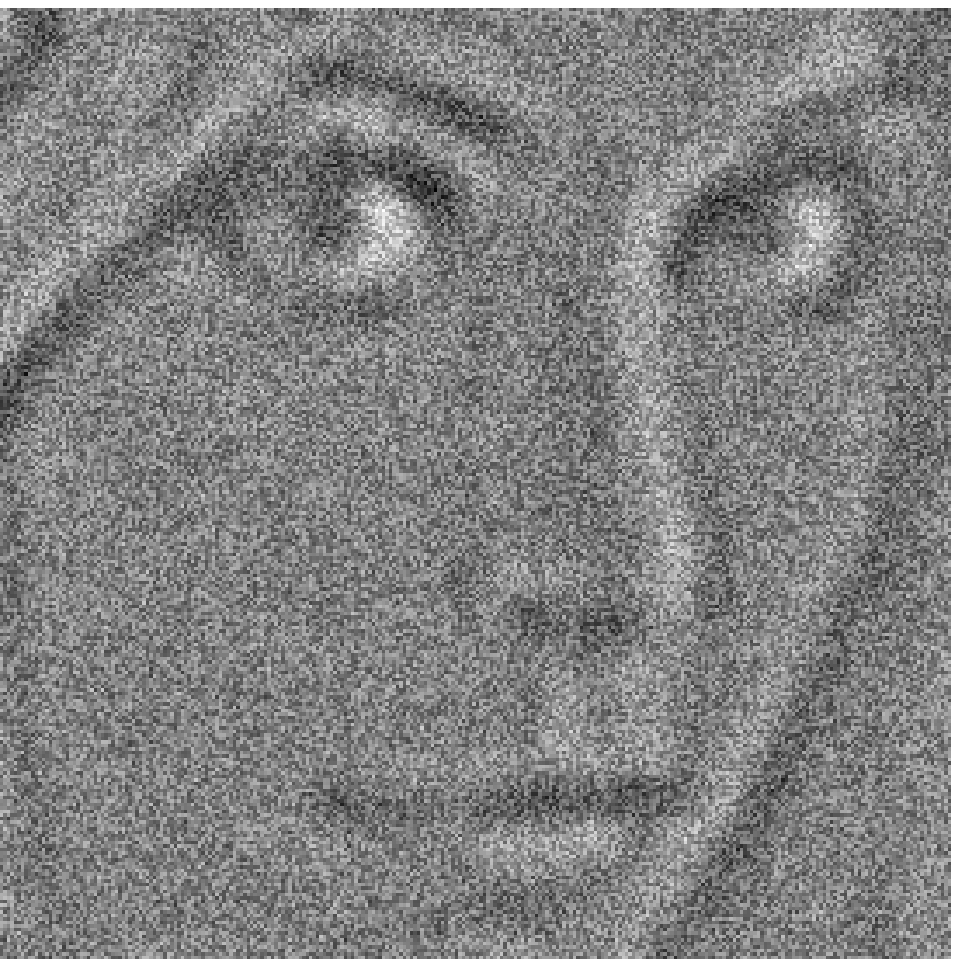} 
} 
} 
\subfigure[]{\parbox{0.45\textwidth}{ 
\epsfxsize=0.45\textwidth 
\epsfysize=0.45\textwidth 
\epsffile{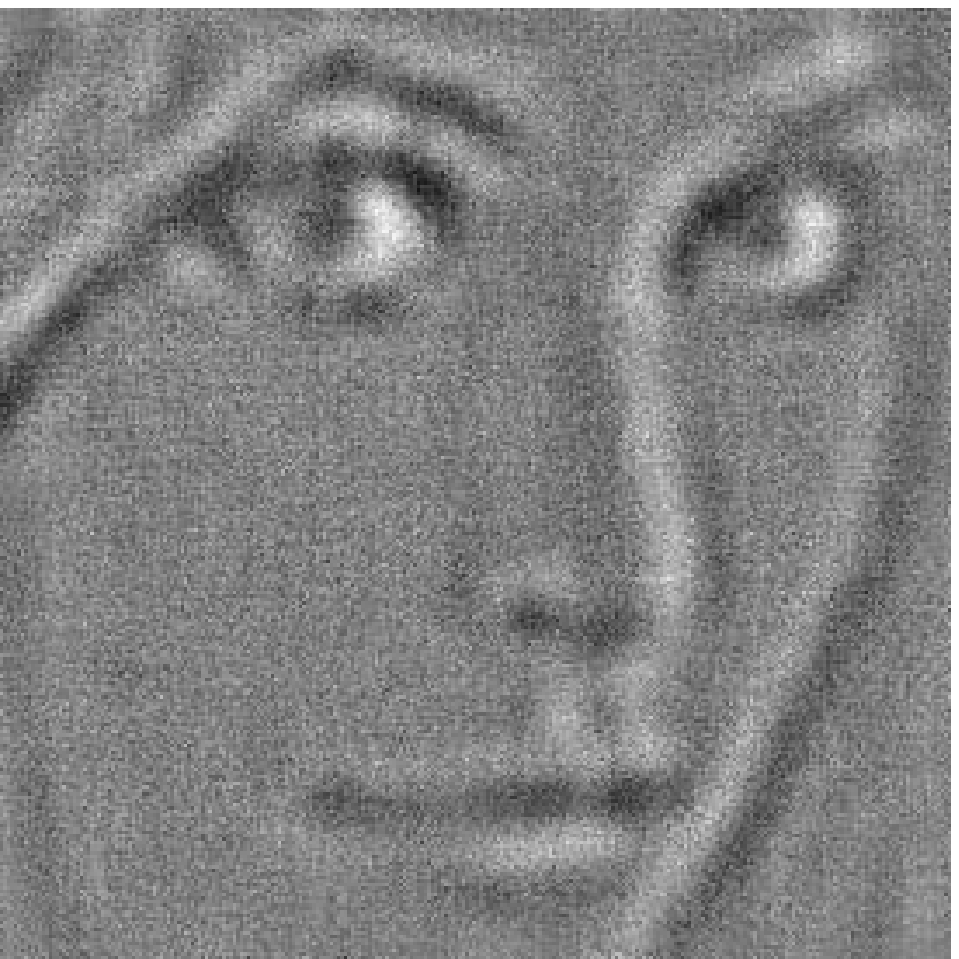}  
} 
} 
\vspace{-0.7cm}\\ 
\end{tabular} 
\end{center}
\vspace{1.1cm}
\caption[]{ a) The LGN image plus noise. To each pixel value has been added 
noise from a flat distribution between -0.1 and 0.1, yielding a signal-to-noise 
ratio of 0.210. b) Reduction in noise level. The image from a) reconstructed 
from the V1 neural activities. The signal to noise ratio is 1.42.} 
\label{fig:p:noise} 
\end{figure}

The summation effect of multiple LGN neurons connecting to each V1 neuron has the property of reducing noise through averaging. The noise considered here would be mixed into the signal along the geniculate axon. The excitation of a V1 neuron is given by
\begin{equation}\label{eq:akeq}
a_k = \sum_j (l_j + \nu_{jk}) \: w_{jk}
\end{equation}
where $\nu_{jk}$ is a random variable of flat distribution between -0.1 and 0.1 for every $j$ and $k$. Let $l^2=\sum_j l_j^2$ and let $\nu^2$ be the variance of the noise distribution.
The squared length of the LGN image divided by the squared length of the noise variance yields a signal to noise ratio of 0.21 for the input. The noise distribution is superimposed on the LGN values in figure \ref{fig:p:noise} a to provide a qualitative illustration of the noise level.

The expected factor $\kappa$ by which the signal to noise ratio should increase from the LGN to V1 is given by:
\begin{equation}\label{eq:sninc}
\kappa = \frac{1}{N}\sum_k \frac{[\sum_j |w_{jk}|]^2}{\sum_j w_{jk}^2}
\end{equation}
where $N$ is the number of V1 neurons and $w_{jk}$ is the connection strength from neuron $j$ of the LGN to neuron $k$ of V1. This equation results from the assumption that the activities in V1 are mainly determined by edges in the input image. It was found that $\kappa=12.01$, slightly less than the value of 16 one would expect for a sum over 16 inputs, because the weights of the inputs are not all identical. 

For the feedback connections
\begin{equation}\label{eq:feedbk}
\Delta l_j = \sum_k a_k \: w_{jk}
\end{equation}
there is again an averaging effect from the summation. By a similar calculation to that for the forward connections, the expected increase in the signal to noise ratio for the backward connections is found to be a factor of 10.66.

For the reconstruction of the image from the integral of V1 activities
\begin{equation}\label{eq:reqeq}
l_j^{recon} = \sum_k A_k \: w_{jk}
\end{equation}
where $A_k$ is the time integral of the activity $a_k$, there is again an averaging effect from the summation. By a similar calculation to that for the forward connections, the expected increase in the signal to noise ratio for the backward connections is found to be a factor of 10.66. Then together, the forward and backward connections yield a factor of 128.03.

A test of the increase in a simulation found a factor relatively close to the predicted value. Twenty feedforward-feedback iterations were applied, with the noise added on the first iteration. The signal to noise ratio of the resulting reconstructed image was 28.39, a factor of 135.2 over the input signal to noise ratio of 0.21.

When the noise is applied on every iteration, the input signal to noise ratio is effectively much higher, because the variance of the total noise added is increased by a factor of twenty. One would predict an output signal to noise ratio of 1.35 in this case. A simulation, whose results are  shown  in figure \ref{fig:p:noise} b, obtained  1.21  for this value. 

The signal-to-noise results are summarized below.

\vskip 20pt
\begin{center}
\begin{tabular}{||r|r|r||}	\hline
Image		  &Computed S/N&Measured S/N\\	\hline\hline
Lena, noise on 1st iteration    &   26.89& 28.39	\\	\hline
Lena, noise on all iterations   &   1.35& 1.21	\\	\hline
\end{tabular}
\end{center}
\vskip 10pt
\begin{center}
\addtocounter{table}{1}Table \arabic{table}\hskip 5pt
{ Signal-to-noise ratios for the lena image.}
\end{center}

\subsection{Entropy Reduction}

The information carrying capacity of the optic nerve is limited, as is that of the  connections between the LGN and V1, and therefore we expect signal compression to be one of  the features of the system. One of the requirements of any information processing system is to utilize the minimum amount of bandwidth to perform its task. Here we will show that the "bandwidth" used by the network for representing the image, defined as the entropy as derived from the activity histograms of the neural layers, decreases as the processing proceeds from one neural area to the next. This indicates that the image is represented more efficiently in the higher layers, as a smaller variety of output activities is needed for the representation. 

\begin{figure}[p] 
\begin{center} 
\begin{tabular}[t]{c} 
\subfigure[]{\parbox{0.45\textwidth}{ 
\epsfxsize=0.45\textwidth 
\epsfysize=0.45\textwidth 
\epsffile{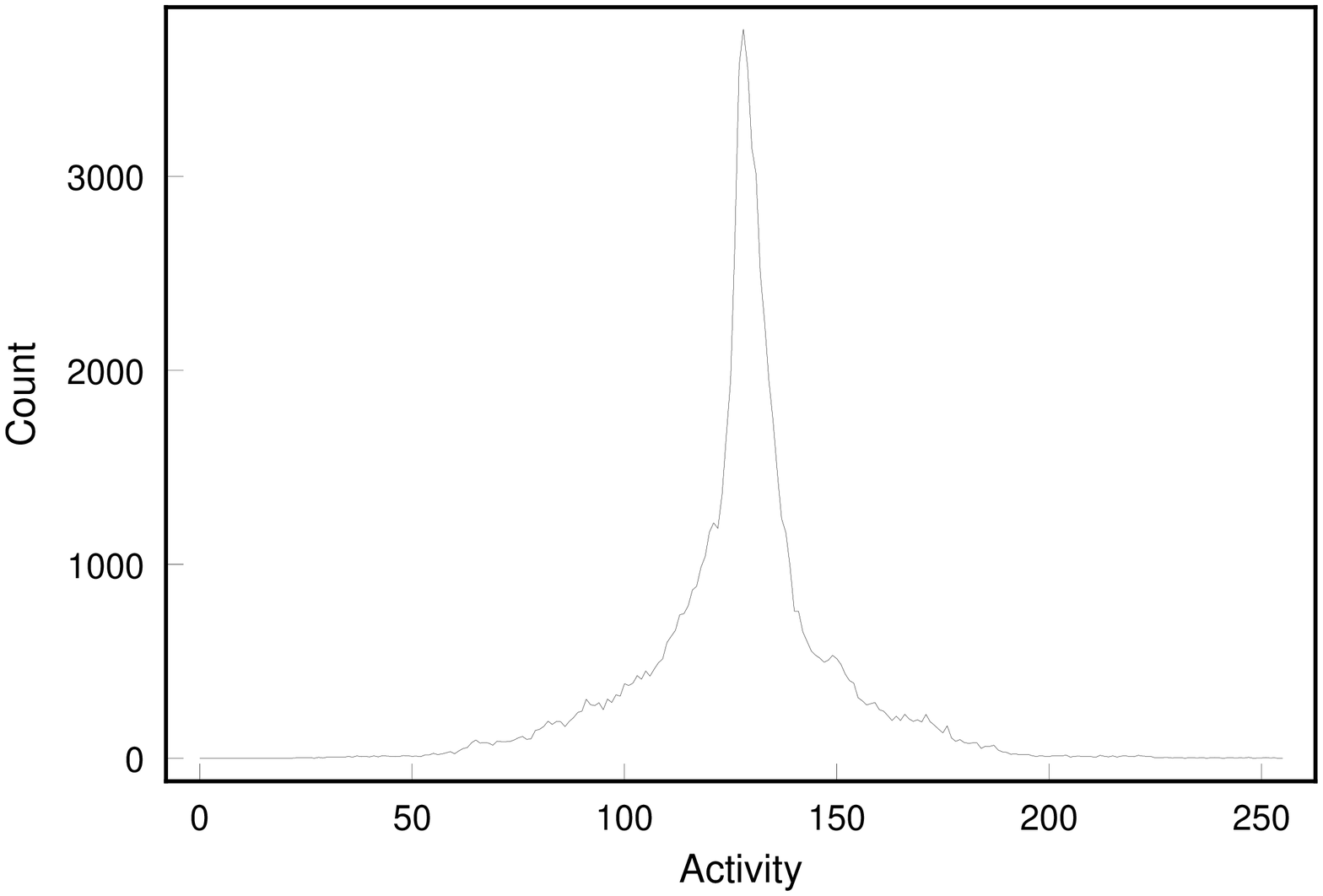} 
} 
} 
\subfigure[]{\parbox{0.45\textwidth}{ 
\epsfxsize=0.45\textwidth 
\epsfysize=0.45\textwidth 
\epsffile{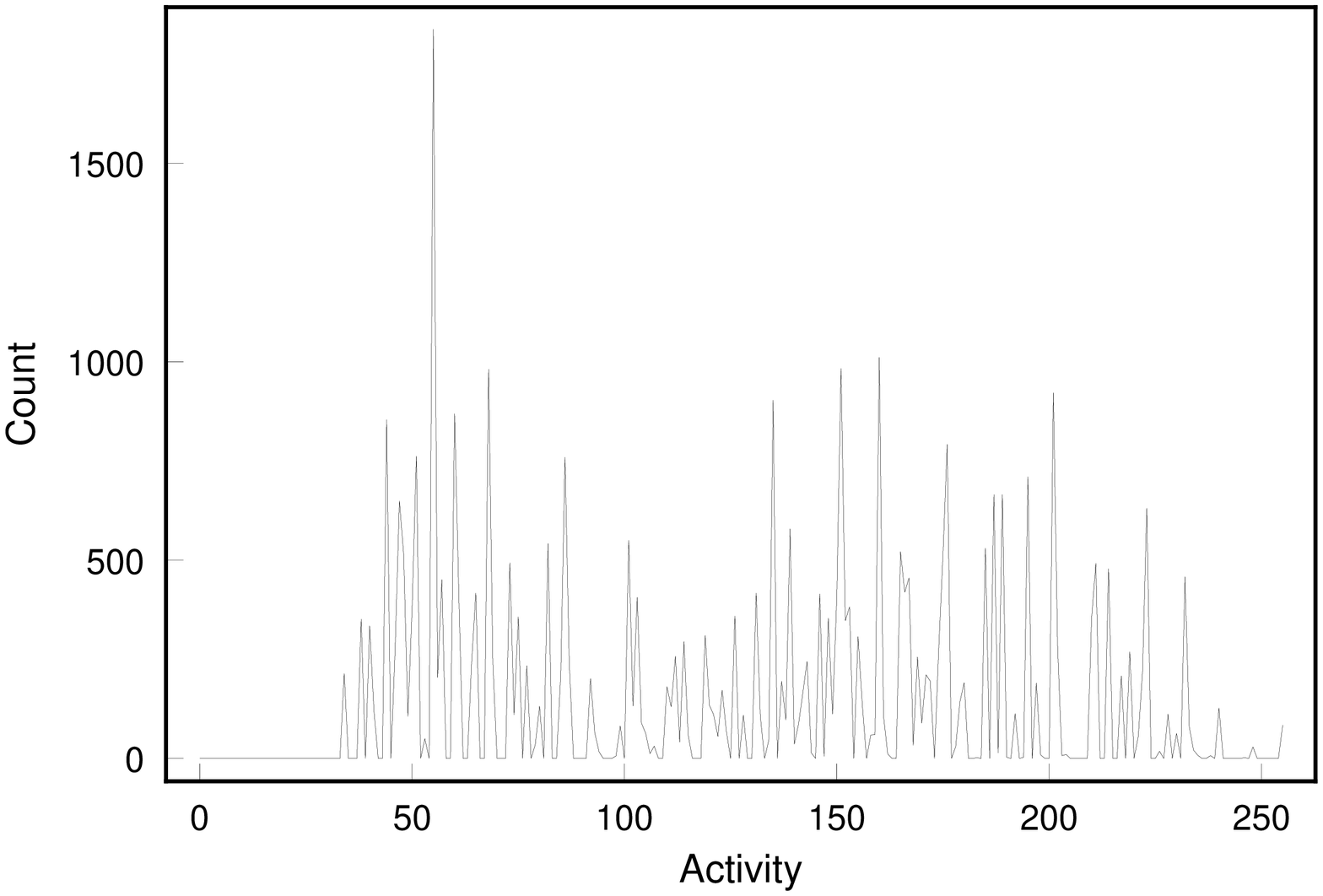}  
} 
} 
\vspace{0.0cm}\\  
\subfigure[]{\parbox{0.45\textwidth}{ 
\epsfxsize=0.45\textwidth 
\epsfysize=0.45\textwidth 
\epsffile{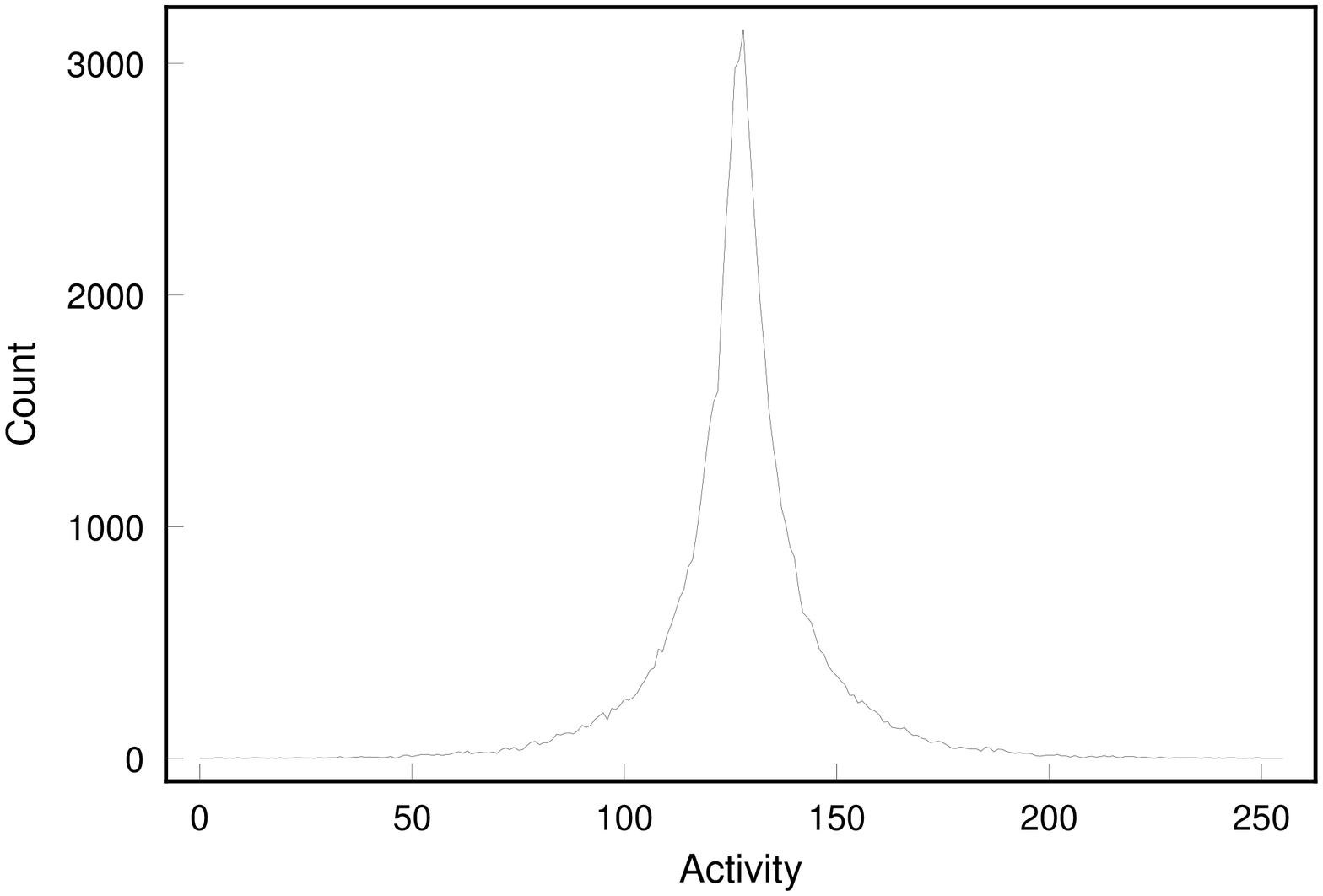}  
} 
} 
\subfigure[]{\parbox{0.45\textwidth}{ 
\epsfxsize=0.45\textwidth 
\epsfysize=0.45\textwidth 
\epsffile{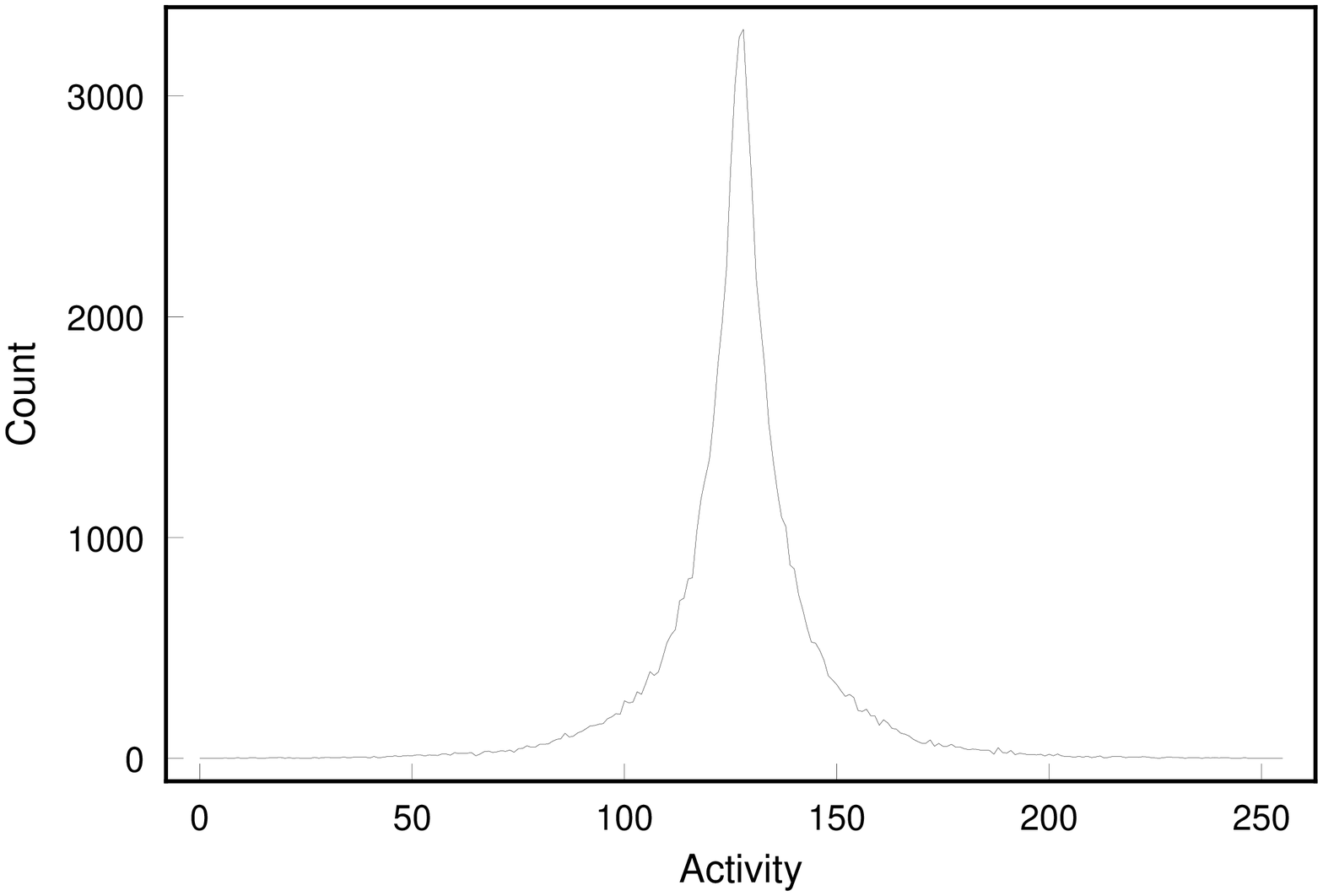}  
} 
} 
\end{tabular} 
\end{center}
\vspace{-0.5cm}
\caption[]{ a) Histogram of activities for original lena image as shown in 
~\protect\ref{fig:p:lena1}a. b) Histogram of input LGN activities for  lena 
image as shown in ~\protect\ref{fig:p:lena1}b. c) Histogram for integral of 
activities of V1 neurons, for lena image. No lateral interaction. d) Histogram 
for integral of activities of V1 neurons, for lena image, lateral interaction 
included.} 
\label{fig:p:hist} 
\end{figure}

Figure \ref{fig:p:hist} a shows the histogram of activities for the original, unconvolved lena image used in the simulations. The horizontal axis goes from 0 to 255 and indicates that the image intensities were encoded into eight-bit numbers. The entropy of the image may be calculated from the histogram. Recall the definition of informational entropy:
\begin{equation}\label{eq:sninc}
I = -\sum_i p_i \log_2(p_i)
\end{equation}
where $p_i$ is the probability of finding a neuron at activity $i$, obtained directly from the histogram of the activities. If the image were completely random, the entropy would be exactly $8$.

Figures \ref{fig:p:hist} b and \ref{fig:p:hist} c show the histograms for the initial LGN activities and the integral of the V1 activities after twenty epochs. The lateral interactions were not included in the simulation. There is a dramatic narrowing of the histogram between the original image and the convolved image input to the LGN, and a slight narrowing from the LGN to V1. With the lateral interactions included, the narrowing from the LGN to V1 is increased, as shown in  \ref{fig:p:hist} d. 
\vskip 20pt
\begin{center}
\begin{tabular}{||l|r||}	\hline
Layer			&	Entropy \\	\hline
Original Image 		& 	7.89 	\\	\hline
Convolved Image 	& 	6.24	\\	\hline
V1 Representation	& 	6.00 	\\	
no lateral interactions	& 	 	\\	\hline
V1 Representation	& 	5.79 	\\	
with lateral interactions 		& 	 	\\	\hline
\end{tabular}
\end{center}
\vskip 10pt
\begin{center}
\addtocounter{table}{1}Table \arabic{table}\hskip 5pt
{ The entropy values for successive stages of processing for the lena image.}
\end{center}

The decreases in entropy may be explained in terms of the statistics of the image being represented and the transformations performed by the network. The convolution is based on the difference-of-gaussians function, which reduces any broad area of constant intensity to near zero. Since the original input image had these kinds of broad areas, the variations in the respective intensities were removed from the image. Likewise, the remaining edges in the LGN image could be detected by V1 neurons whose receptive fields resembled the edges. The activity of a single V1 neuron could then represent the activities of a number of LGN neurons whose activities formed an edge feature. This causes a reduction in image entropy between the LGN and layer V1. The entropy reduction is not more than it is here because image features which are not similar to edges must be represented by relatively complex combinations of edge detectors, which require a greater variety of V1 activities. If the network contained receptive fields of greater complexity, we would expect larger reductions in entropy.

\subsection{Image Compression}

The decrease in entropy described above allows the brain to process images more efficiently, either by decreasing the number of neurons involved in the processing, or by decreasing the required dynamic range of each neuron. Since this gain in efficiency came without any special procedures being added to the network, it is of interest to see whether further gains in efficiency may be possible with some alterations to the  processing. One is in this case seeking not necessarily the natural way of image processing used in the brain, but the most efficient way that can be found using the network described. Nevertheless, it is possible that the algorithms employed may have correlates in the brain's functioning.

As before, the quality of the representation will be analysed from the images reconstructed from the V1 activities. One imagines that one wishes to send an image, sending it in terms of  the integral of the V1 activities. The recipient has knowledge of the weight vectors and understands that the image must be reconstructed with the weight vectors as a basis set and the integral of  the V1 activities as the coefficients. We wish to see if the feature map representation is comparable in performance to other image representation schemes.

\subsubsection{Decreasing the Number of Channels}

Here, a "channel" is one V1 neuron. It is intuitive that if one is able to represent an image using fewer neurons, then one has achieved some compression. An obvious approach here is to send only the highest-valued V1 activities to represent the image. However, this does not work very well. Besides the requirement of sending the coordinates of  the most active neurons, since they change from image to image, the reconstructed image is not good. A cutoff point was chosen below which the coefficient was not included in the reconstruction summation to give a compression ratio of only 17 (the highest one seventeenth coefficients were included), yet the reconstruction was poor.

\subsubsection{Decreasing the Dynamic Range}

The dynamic range is the neural equivalent to the number of  bits per pixel in ordinary image processing. We imagine that the neuron's output is within some finite range, but its value can only be known to a certain accuracy. Therefore, the larger the dynamic range, the more possible values of output that can be discerned. This is comparable to the bits per pixel measure, e.g. eights bits per pixel allows 256 shades to be discerned, whereas four bits per pixel only allows sixteen. 

Image compression can be achieved by decreasing the bits per pixel, since the size of  the representation in bits is the number of channels times the number of bits per pixel, or channel. Therefore decreasing the dynamic range is a form of image compression. In the brain, the dynamic range is limited by the maximum spike rate of a neuron, and gains in terms of this quantity allow more information to be processed within the limitation of the maximum possible spike rate.

\begin{figure}[p] 
\begin{center} 
\begin{tabular}[t]{c} 
\subfigure[]{\parbox{0.45\textwidth}{ 
\epsfxsize=0.45\textwidth 
\epsfysize=0.45\textwidth 
\epsffile{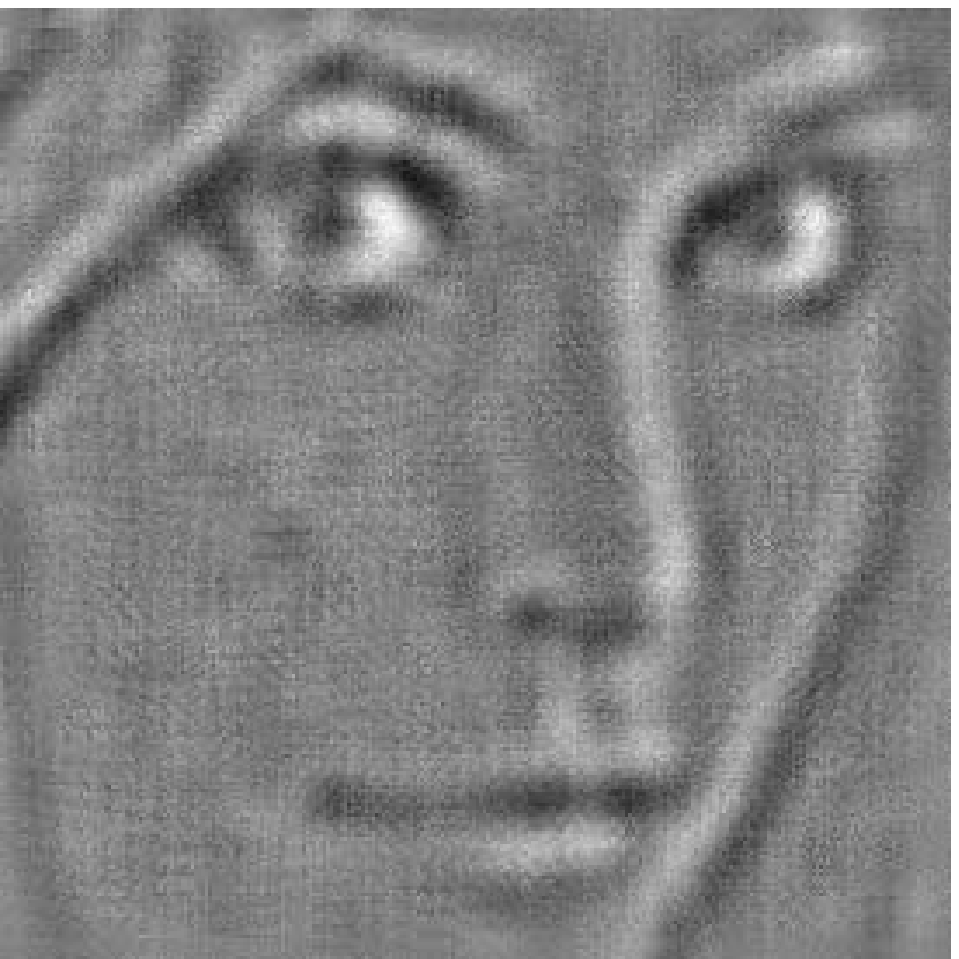 }  
} 
} 
\subfigure[]{\parbox{0.45\textwidth}{ 
\epsfxsize=0.45\textwidth 
\epsfysize=0.45\textwidth 
\epsffile{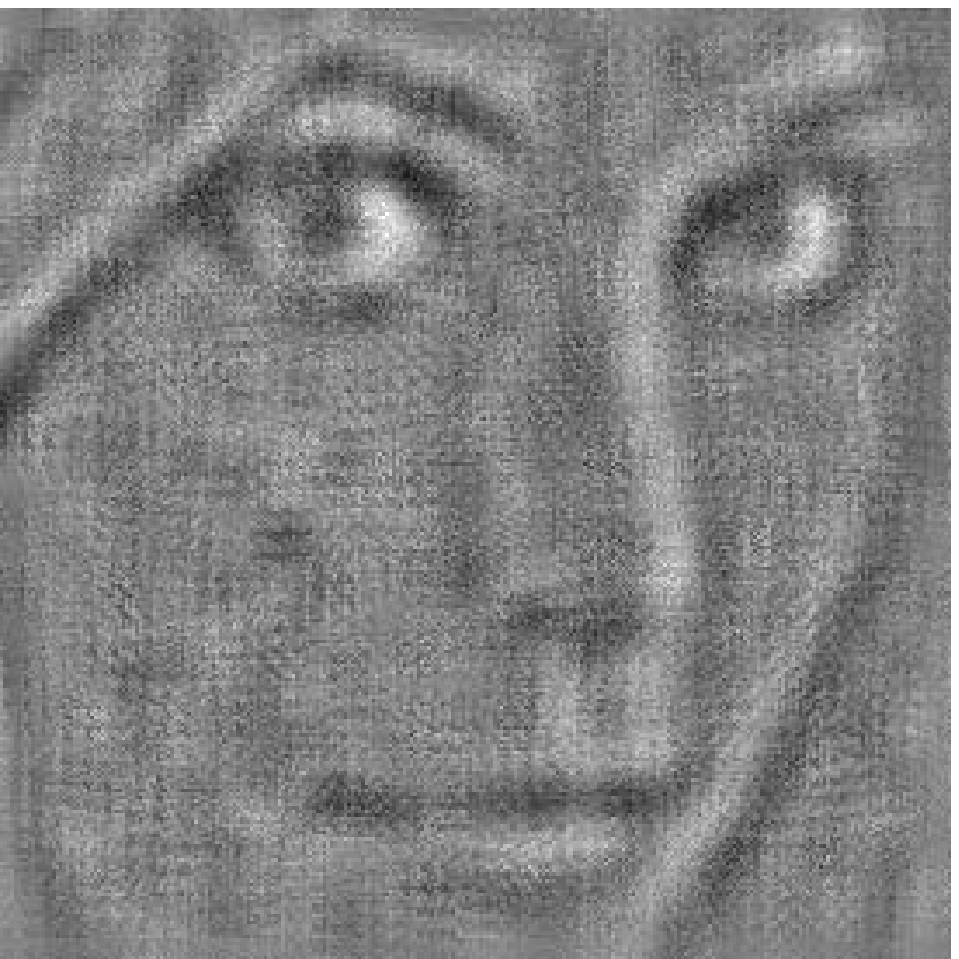 }  
} 
} 
\vspace{0.0cm}\\  
\subfigure[]{\parbox{0.45\textwidth}{ 
\epsfxsize=0.45\textwidth 
\epsfysize=0.45\textwidth 
\epsffile{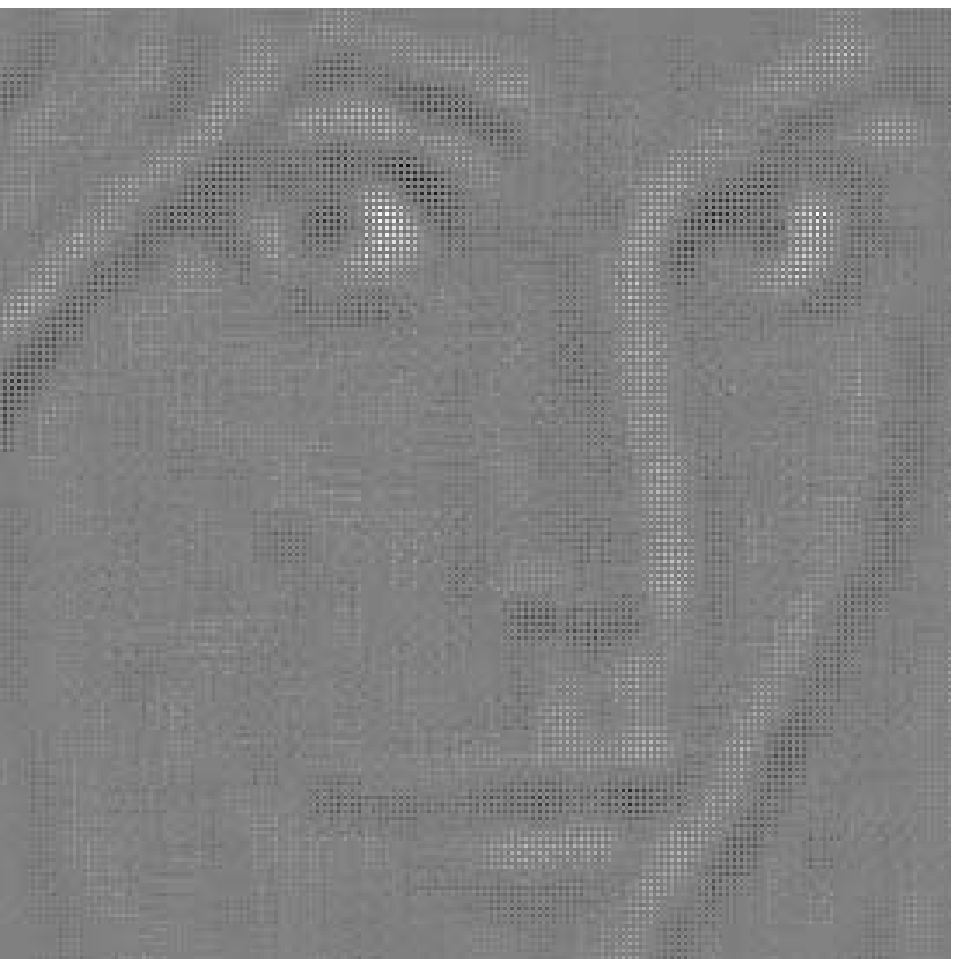 }  
} 
} 
\subfigure[]{\parbox{0.45\textwidth}{ 
\epsfxsize=0.45\textwidth 
\epsfysize=0.45\textwidth 
\epsffile{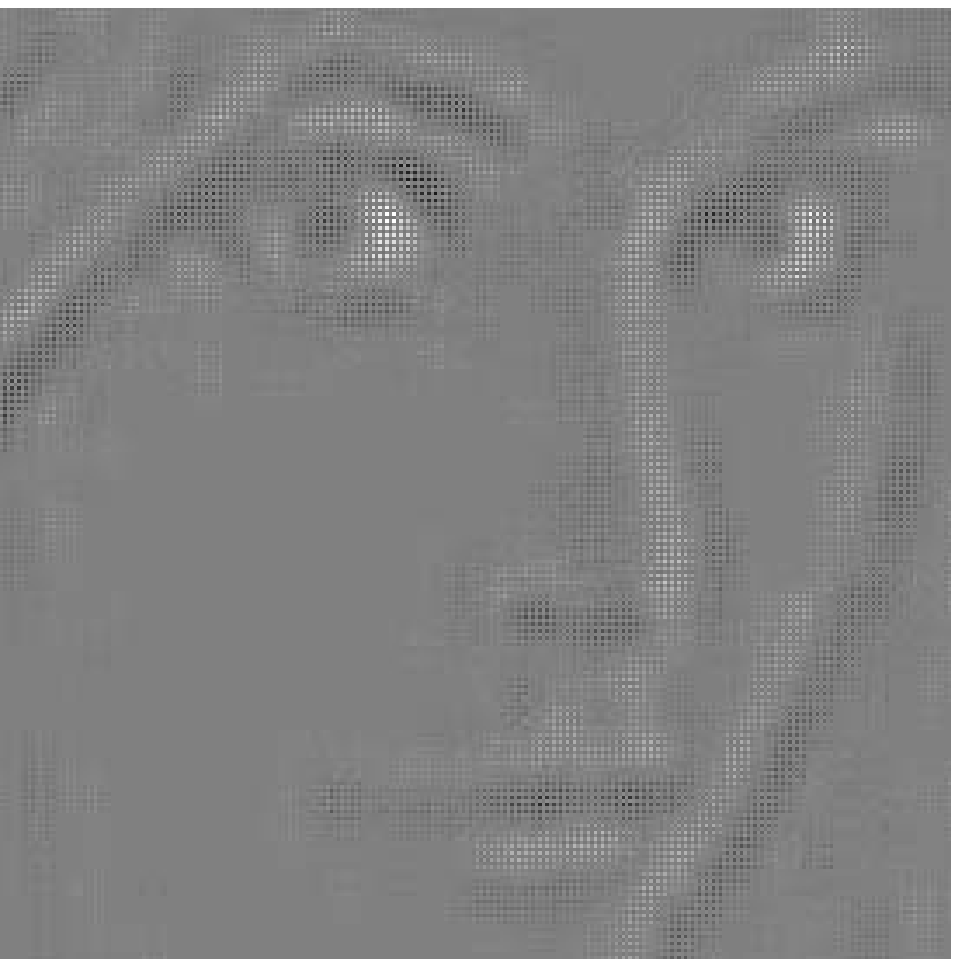}  
} 
} 
\end{tabular} 
\end{center}
\vspace{-1.0cm} 
\caption[]{\footnotesize a) Image compressed in dynamic range by a factor of 16. 
Reconstruction after 20 epochs. b) Image compressed by decreasing dynamic range 
of Integral of V1 activity by a factor of 32. Reconstruction after 20 iterations 
c) Image compressed by a factor of 16 in dynamic range and by a factor of four 
in number of V1 neurons. Total compression 64, or 73.3, taking into account 
difference in image size and V1 size. Reconstruction after 20 iterations. d) 
Image compressed by a factor of 16 in dynamic range, by a factor of four in 
number of V1 neurons, and by a factor of 3.23 in taking only highest valued V1 
neurons. Total compression 206.7, or 236.8, taking into account difference in 
image size and V1 size. Reconstruction after 20 iterations.} 
\label{fig:p:comp} 
\end{figure}

This compression method was tested by converting the integral of the activity of each V1 neuron into an eight-bit number. Truncation of the number by one, two, three, or more bits results in compression factors of two, four, eight, etc. The eight-bit representation was indistinguishable from the representation using floating point numbers. This is not surprising, since the original image was an eight-bit grayscale image. It was found that compression factors of up to sixteen still produced very good reconstructions(figure \ref{fig:p:comp} a), and a compression factor of thirty-two was still acceptable for the simple portrait used(figure \ref{fig:p:comp} b). The iteration of the feedforward-feedback loops seemed to be more important in these cases; the reconstruction converged slowly to the original image as the number of epochs increased.

\subsubsection{Sparse Channels}

Further compression can be obtained by selecting only one-fourth of the  V1 neurons for the reconstruction; every other row and column are excluded. This is combined with the dynamic range reduction to yield a compression ratio of sixty-four to one. The reconstructed images have good quality. See figure \ref{fig:p:comp} c.

\subsubsection{The Three Methods Combined}

The method of choosing the highest-valued pixels can now be added to these two methods. This last method produces a further compression factor of 3.23, for a total compression of 207. The image is of lower quality, but is still recognizable. This is good performance for such a high compression ratio. In fact, when one takes into account the fact that the reconstructed image is $272 \times 272$ while the V1 array is $256 \times 256$ in dimension, a further factor of 1.15 must by applied to all the compression ratios, yielding compressions of  18.4, 73.6, and 237 for the three methods. The results are summarized below.
\vskip 20pt
\begin{center}
\begin{tabular}{||l|r|r||}	\hline
Method	&	Compression Ratio	& Compression Ratio \\
		&				& including array size difference \\ \hline
I. Highest Values of V1 Integrals		& 	17	&	19.6 	\\	\hline
II. Limit Dynamic Range		 	& 	16	 &	18.4	\\	\hline
III. Use Sparse V1 Array		 	& 	  4	 &	  4.6	\\	\hline
IV. Methods II and III			 	& 	64	 &	73.6	\\	\hline
V.  Methods I, II, and III		 	&       207	 &      237	\\	\hline
\end{tabular}
\end{center}
\vskip 10pt
\begin{center}
\addtocounter{table}{1}Table \arabic{table}\hskip 5pt
{ Compression ratios obtained by various methods for the lena image.}
\end{center}

\subsection{Sharpening of Orientational Tuning}

The inhibitory part of the lateral interactions between V1 neurons creates competition between them, for a highly active neuron will suppress the activities of its neighbors. One might expect under this situation that when an edge is presented to the network, the neurons whose receptive fields are best aligned with the edge will win the competition. How does this affect orientational tuning? Without the lateral interaction, a neuron may respond somewhat to an edge that is in its receptive field but not at optimal orientation. With the lateral interaction, such activity will be suppressed by inhibition from neighbors whose receptive fields are more closely aligned with the edge. Only when the orientation of the edge is close to the optimum for the neuron will it retain appreciable activity.

\begin{figure}[p] 
\begin{center} 
\begin{tabular}[t]{c} 
\subfigure[]{\parbox{0.45\textwidth}{ 
\epsfxsize=0.45\textwidth 
\epsfysize=0.45\textwidth 
\epsffile{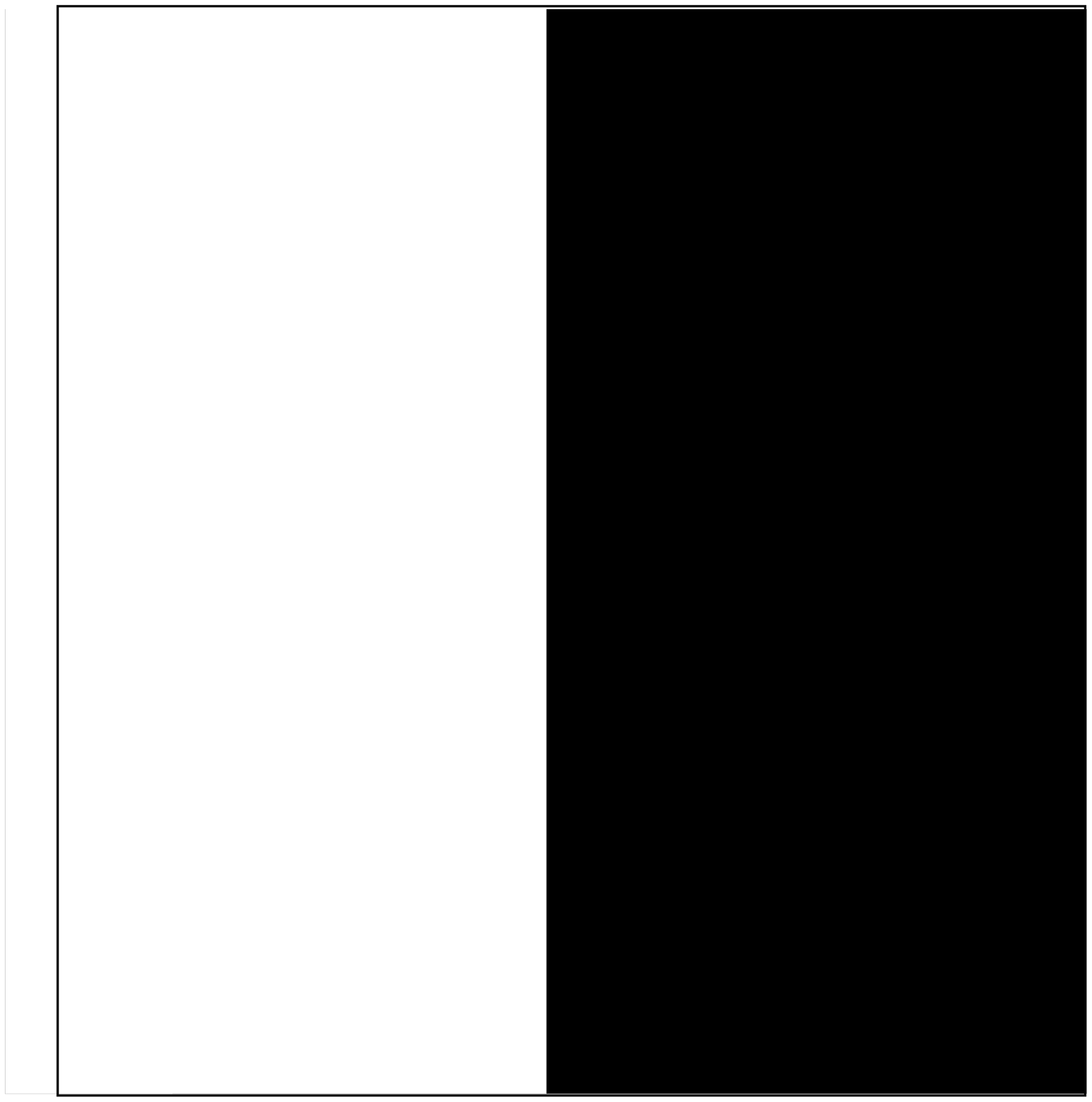}  
} 
} 
\subfigure[]{\parbox{0.45\textwidth}{ 
\epsfxsize=0.45\textwidth 
\epsfysize=0.45\textwidth 
\epsffile{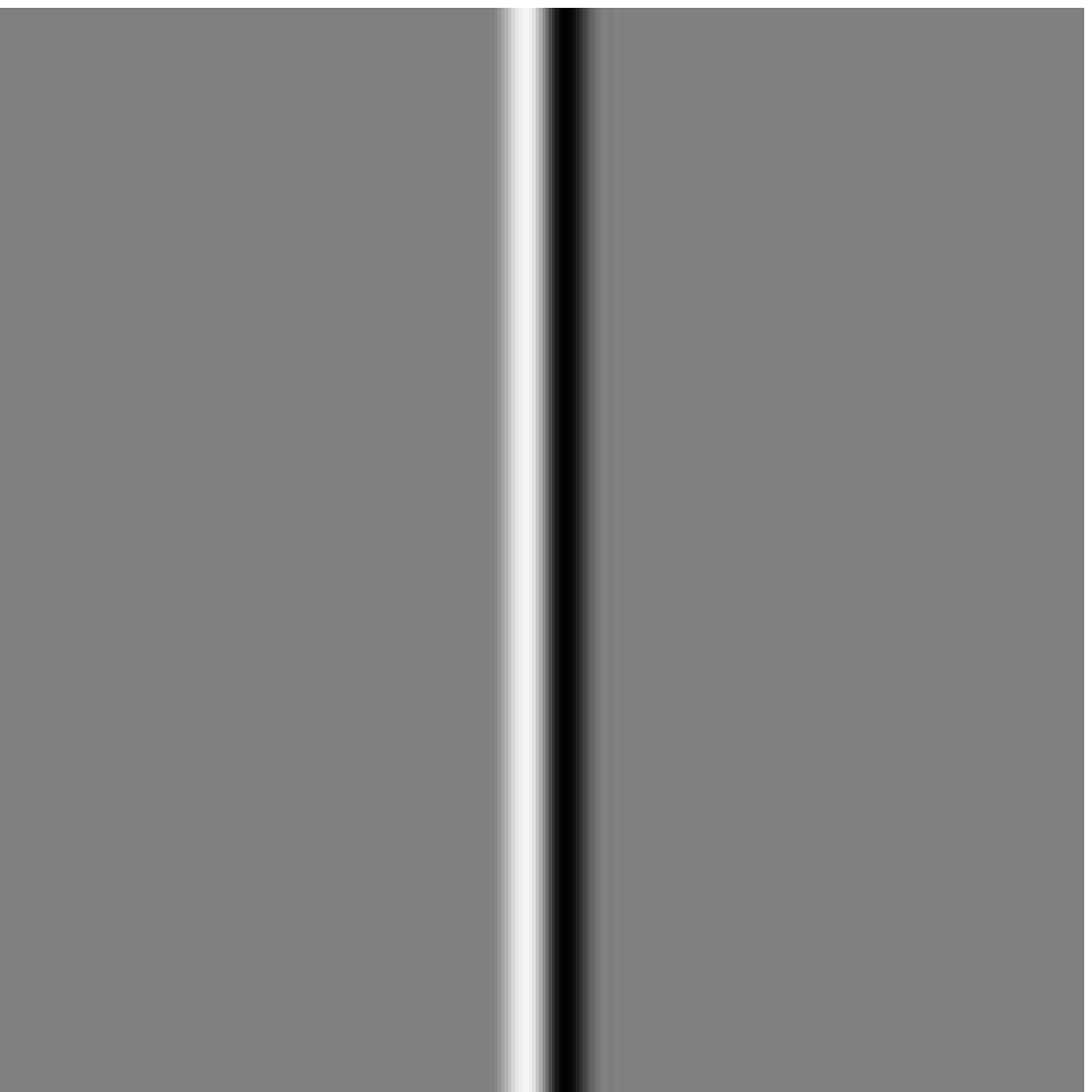}  
} 
} 
\vspace{0.0cm}\\ 
\end{tabular} 
\end{center}
\vspace{-0.5cm}
\caption[]{ a) An original edge image. Surrounding box added for clarity.  b) An 
edge image after convolution. 128 edges  were used, of various orientations 
ranging over one-half revolution. The center of rotation was the center of the 
image.} 
\label{fig:p:edge} 
\end{figure}

\begin{figure}[p] 
\epsfxsize=\hsize 
\epsffile{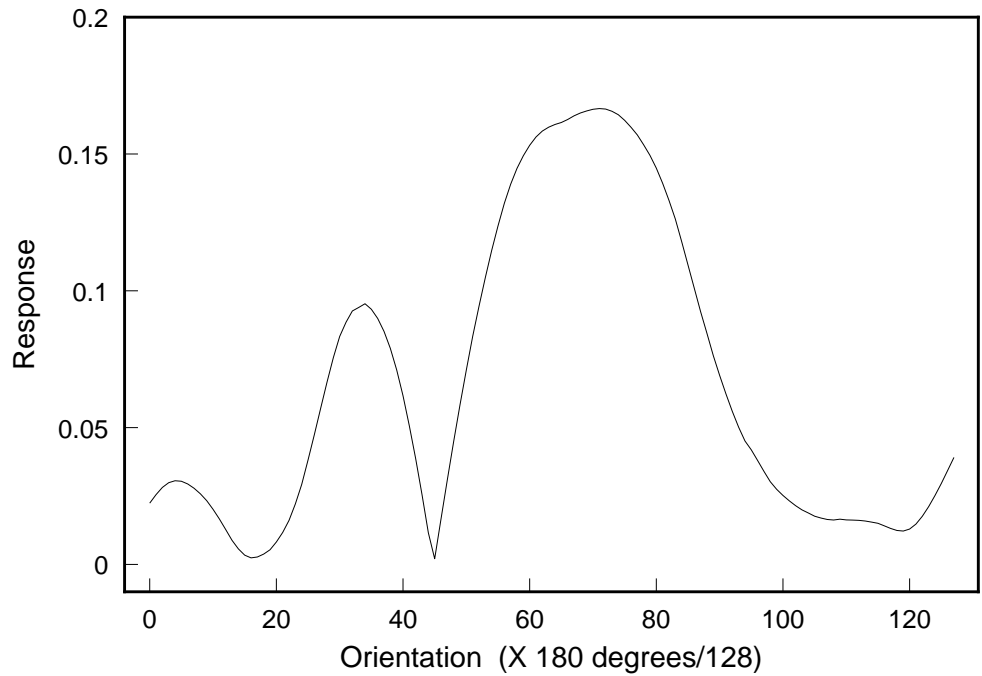}
\vspace{-0.5cm} 
\caption[]{ The sharpening of orientation tuning in V1 is demonstrated by a 
comparison of normalized orientation tuning curves with and without lateral 
interactions bewteen V1 neurons. The figure  shows the absolute value of the 
response of a V1 neuron to a stationary edge of various orientations when no 
lateral interactions are applied. Figure 
~\protect\ref{fig:p:dyn4.16its.128112.abs} shows the response when the lateral 
interactions are iterated sixteen times for each orientation. The range of the 
horizontal axis is one-half of a revolution.} 
\label{fig:p:dyn4.0its.128112.abs} 
\end{figure}

\begin{figure}[p] 
\epsfxsize=\hsize 
\epsffile{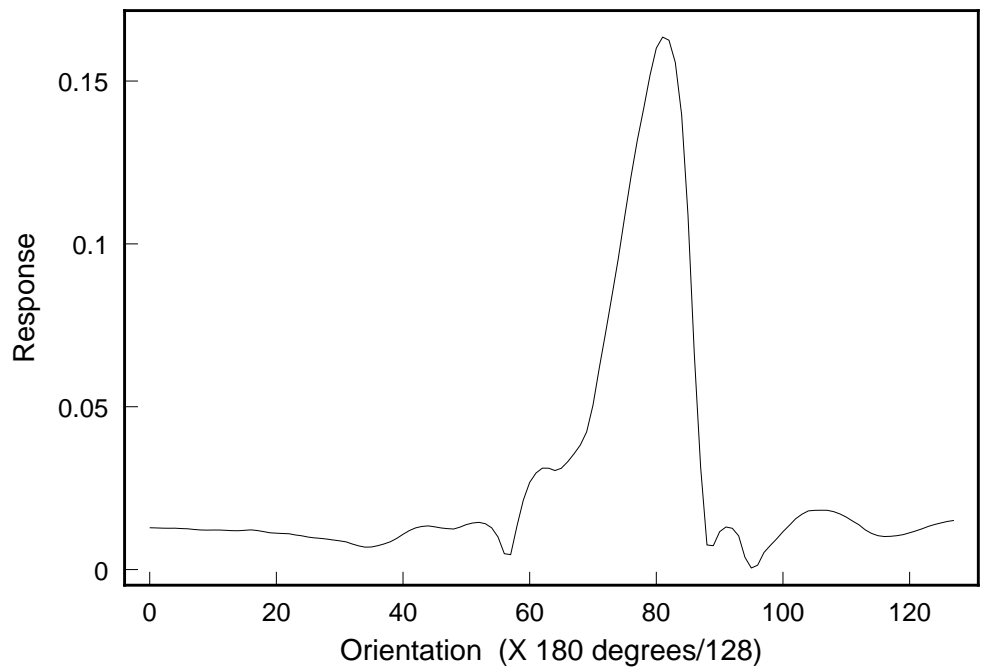}
\vspace{-0.5cm}
\caption[]{The response of the V1 neuron of figure 
~\protect\ref{fig:p:dyn4.0its.128112.abs} to a stationary edge of various 
orientations when the lateral interactions are iterated sixteen times for each 
orientation. A sharpening of the orientational tuning is clearly visible. The 
range of the horizontal axis is one-half revolution.} 
\label{fig:p:dyn4.16its.128112.abs} 
\end{figure}

One hundred twenty-eight edges of various orientations were constructed and convolved through the difference-of-gaussians function. One of the edges and its convolution are shown in figures \ref{fig:p:edge} a and \ref{fig:p:edge} b. The range of orientations was one-half revolution, since the sign of the V1 activities is ignored. The center of rotation was at position (128,128) in V1. The activity of neuron at position (128,112) in V1 was monitored during the simulations. It was found that without the lateral interactions, the neuron exhibited broad orientational tuning, as shown in figure \ref{fig:p:dyn4.0its.128112.abs}.
With the lateral interactions, there is substantial decrease in the width of the orientational tuning curve, as expected. See figure \ref{fig:p:dyn4.16its.128112.abs}.

Others have also proposed mechanisms for the enhancement of orientational tuning. See, for example, ref.\cite{VOGE90A}.

\section{Temporal Aspects of the Image Processing}

The vision system must be responsive to changes in the input, since the organism exists in a dynamic environment. We shall therefore investigate the rate at which the image representation is formed. Since experimental data is often in the form of time series of neural activities, we derive time series results from our simulations.

\subsection{Time Series of V1 Neural Activities}
\begin{figure}[p] 
\epsfxsize=\hsize 
\epsffile{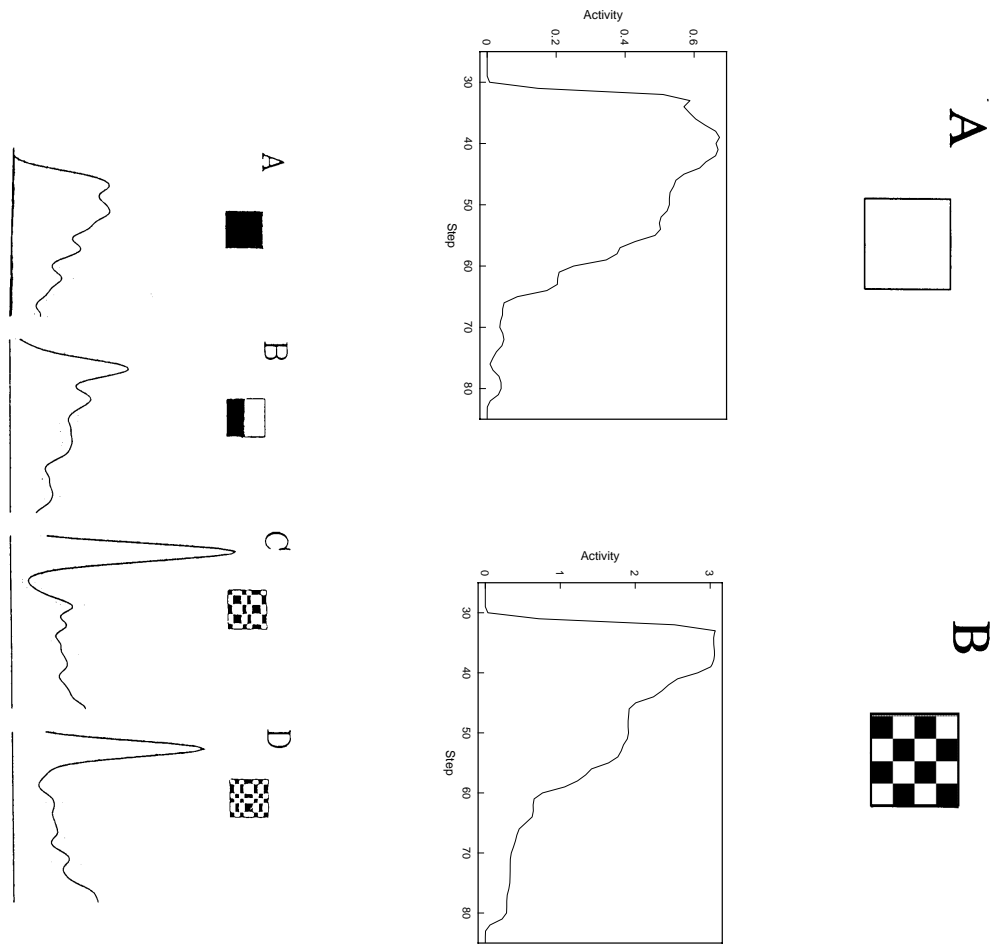}
\vspace{-0.5cm}
\caption[]{ Time series of V1 neuron activity for walsh pattern input. Top: 
Results from simulations. A: Walsh pattern (1,1), B:  Walsh pattern (4,4), 
surrounding box added for clarity, gray background illustrated as white. Middle, 
times series smoothed by gaussian filtering of width two. The curves immediately 
rise upon presentation of an image, then fall as the inhibitory feedback cancels 
the image in the LGN. See figures ~\protect\ref{fig:p:walsh11} and 
~\protect\ref{fig:p:walsh44} for walsh patterns. Bottom: Results and patterns 
from monkey experiments\cite{RICH90A,RICH90B}. } 
\label{fig:p:series} 
\end{figure}

We expect the neural activities in both V1 and the LGN to be transient, for as the inhibitory feedback is applied to the LGN, its activities are damped. When the image has been completely represented by the integral of the V1 activities, the LGN activities have become zero. Then the V1 activities must also be zero, since they depend on feedforward activation from the LGN. Since the V1 activities are assumed to begin at zero, we expect a burst of activity followed by a slower dropoff. This is verified in simulations with walsh patterns, as shown in figure \ref{fig:p:series}.

One can compare these results with similar ones found experimentally for monkeys by Optican and Richmond, et al \cite{RICH90A,RICH90B}, figure \ref{fig:p:series}. These results also show a transient burst of activity. We propose that the transience is due at least in part to inhibitory feedback from V1 to the LGN.

Note that the more complex pattern has a faster falloff in V1 activity. The same result is seen in the simulation. The walsh pattern (4,4) had a faster transient than for pattern (1,1). This was discussed previously in section \ref{walshsec}, as figure \label{fig:p:walsh44} c shows that the image components happen to match the scale of the V1 receptive fields, facilitating the representation and hence the effectiveness of the inhibition. Since the receptive field sizes were carefully scaled in the simulation that created the network to correspond to experimentally-reported sizes, the experimental and simulation results may have the same cause.

\subsection{Time Evolution of the Representation Error}

Since the activity of the LGN after any time step is the difference between the reconstruction and the original image, recall that we took the squared magnitude of the LGN activities as an error measure for the image processing
The error always decreases, and the larger the average magnitude of V1 activities (the larger the signal to V1), the more rapid the decrease in the error of the representation. Hence the use of the eigenvectors of the covariance matrix of the inputs, which maximized the signal.

We would like to compare the measured decrease in the error to the value predicted in equation \ref{eq:dotee}. To do this, we will first measure actual decreases in the error vs time from simulations, then we will find average values of $a_l^2$ from statistical analyses of the images. Since $\dot E$ is not constant in time, we will choose a particular point in time for the measurements, namely, the beginning of the simulation.

\begin{figure}[p] 
\begin{center} 
\begin{tabular}[t]{c} 
\subfigure[]{\parbox{0.45\textwidth}{ 
\epsfxsize=0.45\textwidth 
\epsfysize=0.45\textwidth 
\epsffile{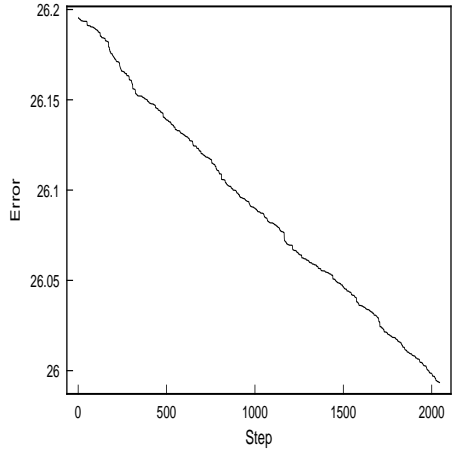 }  
} 
} 
\subfigure[]{\parbox{0.45\textwidth}{ 
\epsfxsize=0.45\textwidth 
\epsfysize=0.45\textwidth 
\epsffile{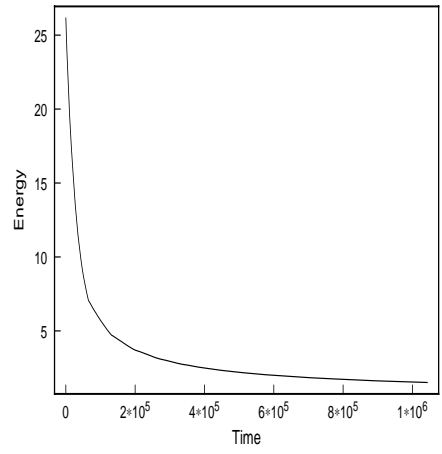 }  
} 
} 
\vspace{0.0cm}\\  
\subfigure[]{\parbox{0.45\textwidth}{ 
\epsfxsize=0.45\textwidth 
\epsfysize=0.45\textwidth 
\epsffile{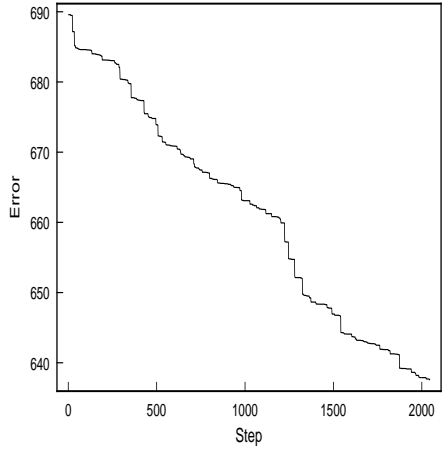 }  
} 
} 
\subfigure[]{\parbox{0.45\textwidth}{ 
\epsfxsize=0.45\textwidth 
\epsfysize=0.45\textwidth 
\epsffile{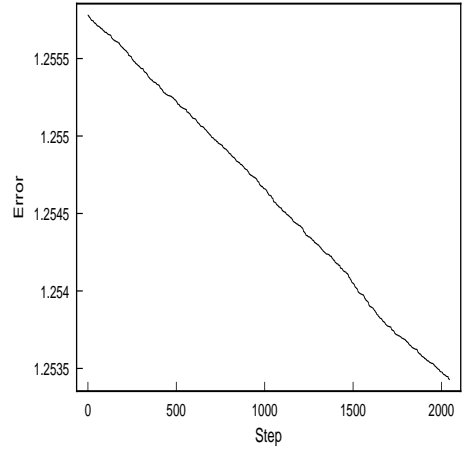}  
} 
} 
\end{tabular} 
\end{center}
\vspace{-0.5cm}
\caption[]{ Initial segment of error vs time, defined as total squared magnitude 
of LGN activity, for the images a)  lena b)  lena (long time) c) walsh pattern 
(1,2) d) random} 
\label{fig:p:energy} 
\end{figure}

The tests were done on three images: the lena image, the walsh pattern (1,2) image, and a random image(random noise of flat distribution in the retina, convolved as usual through the difference-of-gaussians to produce the input to the LGN). A factor of $0.1$ was inserted into equation \ref{eq:inhib} to make the change in the error more gradual so that the approximation of the change in the error as a derivative in eq. \ref{eq:dotee} would be valid. This factor was compensated for in subsequent calculations. After each time step, the squared magnitude of the LGN activities was computed. Recall that the activity values are updated by the choosing of one V1 neuron at a time and that this defines one time step. The resulting plots of error vs time are shown in figure \ref{fig:p:energy}. Figure \ref{fig:p:energy} b shows the long time result for the lena image. Note that the curve is slower than exponential, showing that at later times the proportion of  $\langle a_l^2 \rangle$ to $E$ decreases. As the processing progresses, the image remaining in the LGN contains less components that are matched to the edge detectors, because these components have been subtracted from the image by the inhibition. The remaining components are of higher order than the simple edge detectors comprising the weight vectors and do not cause as high values of $\langle a_l^2 \rangle$ in proportion to $\langle l_j^2 \rangle$.

\subsection{Average Values for Neural Activities}

The value of $\langle \dot E \rangle$ is predicted by the dynamical equations to be $-\langle a_k^2 \rangle$. For this comparison, the values of  $a_k$ were calculated by applying the weight vectors to the input LGN image. The image was not altered by any backward inhibition, since the time period of interest is the beginning of the simulation, when the effect of the overall LGN image is negligible. Means and variances for $\langle a_k^2 \rangle$, $\langle l_j^2 \rangle$, $\langle \sum_j w_{jk}^2 \rangle$, and $\langle \lambda_k^2 \rangle$ were computed in this way from 65,
65,536 samples. The data, and comparisons to the predicted values, are summarized in the tables presented below.
\vskip 20pt
\begin{center}
\begin{tabular}{||r|r|r||}	\hline
	Lena 		& Mean		& Variance\\	\hline
V1 activity squared	& 9.89e-04	& 6.20e-06\\	\hline
LGN activity squared	& 9.26e-04	& 3.34e-06\\	\hline
Eigenvalue		& 2.28e-03	& 3.16e-05\\	\hline
\end{tabular}
\vskip 20pt

\begin{tabular}{||r|r|r||}	\hline
Walsh Pattern (1,2)	& Mean		& Variance\\	\hline
V1 activity squared	& 2.93e-02	& 3.29e-02\\	\hline
LGN activity squared	& 2.10e-02	& 9.65e-03\\	\hline
Eigenvalue		& 6.63e-02	& 1.62e-01\\	\hline
\end{tabular}

\vskip 20pt

\begin{tabular}{||r|r|r||}	\hline
	Random Image	& Mean		& Variance\\	\hline
V1 activity squared	& 1.19e-05	& 3.02e-10\\	\hline
LGN activity squared	& 3.82e-05	& 2.53e-09\\	\hline
Eigenvalue		& 2.75e-05	& 1.56e-09\\	\hline
\end{tabular}
\end{center}
\vskip 10pt
\begin{center}
Tables 	
$\addtocounter{table}{1}		\arabic{table},
\addtocounter{table}{1} 	\arabic{table},
\addtocounter{table}{1} 	\arabic{table}$\hskip 5pt
{ Means and variance of network quantities for various images.}
\end{center}

\vskip 20pt
\eject
\begin{center}
\begin{tabular}{||r|r|r|r||}	\hline
Image	&$\langle \dot E \rangle$&$-\langle a_k^2 \rangle$&Ratio\\\hline\hline
Lena 		   &	-9.33e-04 &  9.49e-04& 1.017\\	\hline
Walsh Pattern (1,2)&	-2.52e-02 &  2.93e-02& 1.16\\	\hline
Random Image	   &	-1.15e-05 &  1.12e-05& 0.974\\	\hline
\end{tabular}
\end{center}
\begin{center}
\addtocounter{table}{1}Table \arabic{table}\hskip 5pt
{ Comparisons of error slopes measured from dynamical simulations and predicted from statistical averages of V1 activities for various images.}
\end{center}
The average rate of change of the error at the beginning of the dynamics simulation as predicted from statistical measures of the V1 activity is close to the measured rate, except for the walsh pattern. Being as simple as it is, the walsh pattern would certainly have the most extreme statistics of the three images. One sees from the plot of error vs. time for the walsh pattern (figure \ref{fig:p:energy} c) that the time evolution of the error is much more uneven here than in the other two cases. A V1 activity, and therefore a change in the error is either high or low for this image, because it is composed only of edges or blank spaces. The approximation for the change in the error as a derivative in \ref{eq:dotee} may not be valid for such changes.

\subsection{The Role of the Eigenvalues of the Covariance Matrix}

Let us now examine how to predict the behavior of the network from just the knowledge of the eigenvalues of the covariance matrix of the inputs, since this allows one to make contact with various computational schemes involving eigenvalues. Define a "weight vector" $w_{k}$ as a vector composed of values running over the $j$ index of $w_{jk}$. The receptive field of neuron $k$ in V1 as defined by the set $w_{jk}$ may be interpreted as a weight vector $w_{k}$. With the learning algorithm presented, the weight vectors develop into eigenvectors of the covariance matrix. Consider the rate of decrease of the error
\begin{eqnarray}\label{eq:lambda}
\langle \dot E \rangle &=&  \langle \sum_{j }\:  y_j \dot{y}_j \rangle \\
\nonumber\\
&=&  \langle -\sum_{j} \:  y_j  \: a_l \: w_{jk} \rangle\\
\nonumber\\
&=&  \langle -\sum_{ij} \:  y_i  \: y_j   \: w_{ik} \: w_{jk} \rangle\\
\nonumber\\
&=&   -\sum_{ij} \:  \langle y_i  \: y_j \rangle  \: w_{ik} \: w_{jk} \\
\nonumber\\
&=&   -\sum_{ij} \:  R_{ij}  \: w_{ik} \: w_{jk} \\
\nonumber\\
&=&   -\sum_{j} \:  \lambda_k  \: w_{jk} \: w_{jk} \\
\nonumber\\
&=&   -  \lambda_k  \: |w_{k}|^2 \\
\nonumber
\end{eqnarray}
where $R_{ij}\equiv \langle y_i  \: y_j \rangle$ and $\lambda_k$ is the eigenvalue of $w_{k}$. Therefore the speed of the network depends on the magnitude of the eigenvalues. By symmetry, the value of $\lambda_k$ is in principle independent of $k$. Eigenvalues were found by dividing $-\langle a_k^2 \rangle (=-\dot E)$ by $|w_{jk}|^2$ for various $k$. The image used to find the $a_k$ values was the random image, since that was the image used to develop the weights and is expected to have the most uniform statistics. Measurements of the weight values of the network show that the mean and variance of $\lambda_k$ are $2.75 \times 10^{-5}$ and $1.57 \times 10^{-9}$, respectively.

\subsubsection{Testing for Eigen Properties}

\begin{figure}[p] 
\begin{center} 
\begin{tabular}[t]{c} 
\subfigure[]{\parbox{0.45\textwidth}{ 
\epsfxsize=0.45\textwidth 
\epsfysize=0.45\textwidth 
\epsffile{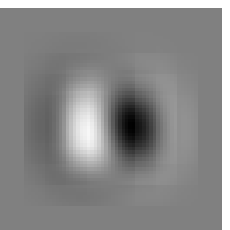}  
} 
} 
\subfigure[]{\parbox{0.45\textwidth}{ 
\epsfxsize=0.45\textwidth 
\epsfysize=0.45\textwidth 
\epsffile{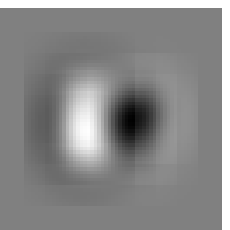}  
} 
} 
\vspace{0.0cm}\\ 
\end{tabular} 
\end{center}
\vspace{-0.5cm}
\caption[]{ a)   Weight vector used to check for eigen property. b) Average of 
product of the weight vector and lgn activities, random inputs. The resemblance 
to the original weight vector shows that the weight vector is an eigenvector.} 
\label{fig:p:eigfind4.200.weight} 
\end{figure}

We should now like to verify whether the receptive fields $w_k$ obtained in the simulations of cortical development are really the eigenvectors of the covariance matrix of the inputs. We cannot construct the covariance matrix directly; it is too large, but we can approximate the covariance matrix by sequential averaging over a series of inputs. Let $R_ij$ be the covariance matrix of inputs $y_i$ and  consider
\begin{eqnarray}\label{eq:aveprod}
\langle \sum_j y_i \: y_j \: w_{jk} \rangle &=& \sum_j R_{ij} \: w_{jk}\\
\nonumber\\
&=& \lambda_k \: w_{ik}\\
\nonumber
\end{eqnarray}
if $w_k$ is an eigenvector of $R$, where $\lambda_k$ is its eigenvalue. Since
\begin{equation}\label{eq:defa}
\sum_j y_i \: y_j \: w_{jk}  = y_i \: a_k
\end{equation}
we must determine whether
\begin{equation}\label{eq:avela}
\langle y_i \: a_k \rangle =  \lambda_k \: w_{ik}
\end{equation}
for some $\lambda_k$. The average is easy to find by applying the known weight values to a series of images. A set of weights  $w_{jk}$ was chosen arbitrarily and a series of 200 random images was presented. Figure \ref{fig:p:eigfind4.200.weight} a shows the weight vector chosen and \ref{fig:p:eigfind4.200.weight} b shows the quantity $\langle y_i \: a_k \rangle$. It is plain to see that the product is a scalar multiple of the weight vector, and hence the weight vector is an eigenvector of the covariance matrix. The closeness of the product to the weight vector was measured as the cosine of the angle between the two, e.g., their inner product divided by the square roots of their magnitude. This value was found to be 0.949. The eigenvalue was taken as the ratio of their norms and was $2.87 \times 10^{-5}$.
This is not far from the average value obtained by dividing $\langle a_k^2 \rangle$ by $|w_{jk}|^2$ for various $k$.

\subsubsection{Comparing the Dynamics Results to the Eigenvalue Prediction}

The mean and variance of $|w_{k}|^2$ were measured directly and found to be $4.33 \times 10^{-1}$ and $3.71 \times 10^{-3}$, respectively. Thus the value of $\langle \dot E \rangle$ may be predicted from the known eigenvalues \cite{CUN91A}. The table below compares the predicted and measured values.

\vskip 20pt
\begin{center}
\begin{tabular}{||r|r|r|r||}	\hline
Image	&$10\langle \dot E \rangle$&$\langle \lambda_{k} |w_{k}|^2\rangle$&Ratio\\\hline\hline
Lena 		   &	-9.33e-04 & 9.37e-04 & 1.004\\	\hline
Walsh Pattern (1,2)&	-2.52e-02 & 2.85e-02 & 1.13\\	\hline
Random Image	   &	-1.15e-06 & 1.12e-05 & 0.974\\	\hline
\end{tabular}
\end{center}
\begin{center}
\addtocounter{table}{1}Table \arabic{table}\hskip 5pt
{ Comparisons of error slopes measured from dynamical simulations and predicted from eigenvalues for various images.}
\end{center}
The predictions again close to the actual values.

\section{The Fokker-Planck Equation for the Error Value of the Image Processing Network}

To construct a Fokker-Planck equation for the error, $E$, we must find its first and second moments while the network is processing the image. The first moment is clearly $-\langle a_k^2 \rangle$. For the second moment, we have
\begin{eqnarray}\label{eq:dele}
 \frac{\langle(\Delta E)^{2}\rangle}{\Delta t} &=&\frac{\langle(E(t+\Delta t)-E(t))^{2}\rangle}{\Delta t}\\
\nonumber\\
&=&2\langle\dot{E}(t)(E(t+\Delta t)-E(t))\rangle\\
\nonumber\\
&=&2\langle\dot{E}(t)\int_{t}^{t+\Delta t} dt' \dot{E}(t')\rangle\\
\nonumber\\
&=&2\langle\langle \dot{E} \rangle\int_{t}^{t+\Delta t} dt' \dot{E}(t')\rangle\\
\nonumber
\end{eqnarray}
Since $t$ is arbitrary in the above equation, we can replace $\dot{E}(t)$ by
$\langle \dot{E} \rangle$.

In the dynamics simulations, $l$ for a given time step is chosen randomly. The dynamics as implemented is therefore a Langevin equation for the motion of $E$:
\begin{equation}\label{eq:lang}
\dot E  = -a_{l(t)}^2
\end{equation}
Now, $-a_{l(t)}^2$ is considered as an impulse function, since time is discrete in the simulation. Thus the integral in eq. \ref{eq:dele} can be replaced by the value of $\dot E$ at some time other than $t$. This yields
\begin{eqnarray}\label{eq:dele2}
 \frac{\langle(\Delta E)^{2}\rangle}{\Delta t} &=&2\langle a_l^2 \rangle^2\\
\nonumber
\end{eqnarray}
Let $\alpha=\langle a_l^2\rangle$. The Fokker-Planck equation is \cite{STRA89A}
\begin{equation}\label{eq:fpe}
\frac{\partial \sigma(E,t)}{\partial t} =
\frac{\partial}{\partial E}\lbrace \alpha-\frac{\partial}{\partial E}\alpha^{2}\rbrace\sigma(E,t)
\end{equation}
For the short times considered here, we can neglect any dependence of  $\alpha$ on $E$. To solve the equation, we note that the probability distribution for the error function is a delta function, since the initial error is known as exactly the squared length of the input LGN image. Thus the solution is
\begin{equation}\label{eq:lang}
\sigma(E,t) =
\frac{1}{2\alpha\sqrt{\pi t}} \exp{\frac{-(E-E(t))^2}{4\alpha^2t}}
\end{equation}
where $E(t)$ is the expectation value of $E$ at time $t$, according to the initial value of $E$ and the expectation value of $\dot E$. In the following, we will compare the results of the dynamical simulations to the prediction of the Fokker-Planck equation.

\subsection{Variance Measurements from the Dynamics Simulations}

\begin{figure}[p] 
\epsfxsize=\hsize 
\epsffile{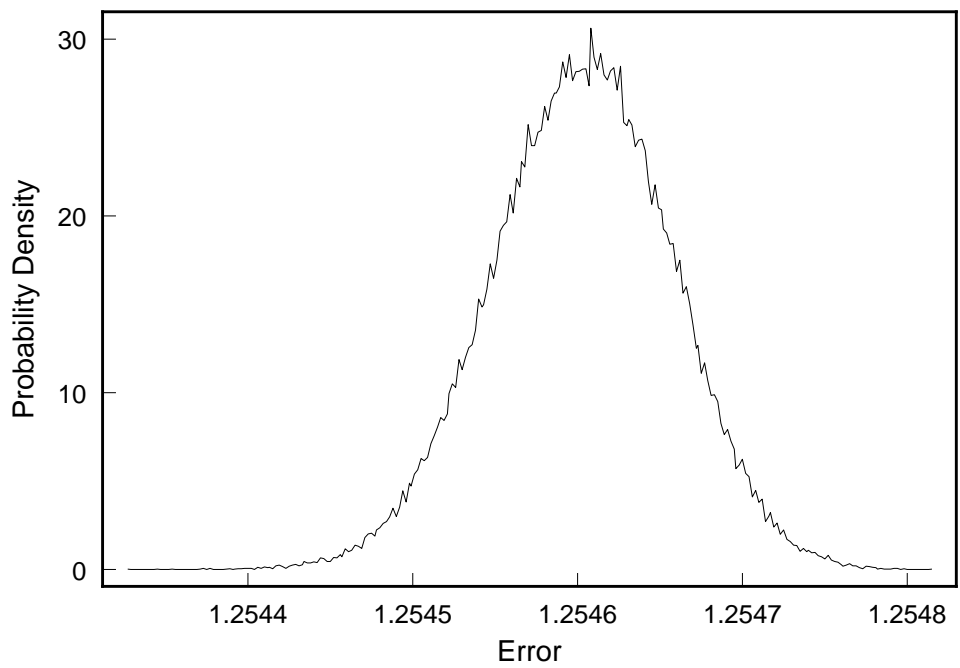}
\vspace{-0.5cm}
\caption[]{ The histogram of error values at $t=1024$ for one hundred thousand 
runs. The simulation was done on the random image. Note the gaussian shape of 
the curve, showing that the solution to the Fokker-Planck equation is of the 
correct form.} 
\label{fig:p:energygaussr} 
\end{figure}

The slopes were measured from the plots of $E$ vs $t$ in figure \ref{fig:p:energy} to find the expectation value of $\dot E$ at $t=0$. The values are presented in the table below. Because the order of V1 neurons chosen is random, there will be some variance in the value of the error at a given time. To find  the shape of the probability function for the error, the first one thousand and twenty-four steps of the image processing was repeated one hundred thousand times with different choices of the V1 neurons. Then a histogram was made from the one hundred thousand values of $E$ at $t=1024$. The histogram is plotted in figure \ref{fig:p:energygaussr} and shows the probability density of $E$ at that time. Note the gaussian shape of the histogram, which shows that the form of the Fokker-Planck solution is correct. The variances were measured numerically and are included in the table below.

Let us check the solution to the Fokker-Planck equation. The variance is given by
\begin{equation}\label{eq:var}
\sigma = 2 \: \alpha^2 \: t
\end{equation}
$-\alpha$ is the average slope of $E$ vs $t$. The values of $\alpha$ obtained from the plots of $E$ vs $t$ are used to calculate the expected variance of the error at $t=1024$, and this is compared to the measured variance in the following table.

\vskip 20pt
\begin{center}
\begin{tabular}{||r|r|r|r||}	\hline
Image		  &$\langle \dot E \rangle$&Computed Variance&Measured Variance\\	\hline\hline
Lena 		   &	-9.49e-04 & 1.84e-03 & 5.14e-03	\\	\hline
Walsh Pattern (1,2)&	-2.93e-02 & 1.76e+00 & 3.04e+01	\\	\hline
Random Image	   &	-1.12e-05 & 2.56e-07 & 2.97e-07	\\	\hline
\end{tabular}
\end{center}
\vskip 10pt
\begin{center}
\addtocounter{table}{1}Table \arabic{table}\hskip 5pt
{ Computed and measured variances for the error of the representation at $t=1024$.}
\end{center}

For the random image, the actual variance is within reasonable agreement to the predicted variance. For the others, the prediction is quite far off. This is presumably due to the distribution of $a_l^2$ being somewhat bimodal; some high values where edges occur in the images, and low values where there are no edges. This would no doubt increase the variance of the error over that predicted, and indeed one sees a greater error for the walsh function, where this bimodality is more pronounced.

\section{Discussion}

The simulations show that the system for representing images proposed has several desirable properties in regards to efficiency and robustness to noise. The scheme is suitable for compressing images at very high ratios.  
We also found that the time series of V1 neuron activities resembles those found in experiments. 

Because of the inhibitory feedback, we could define an error as the squared length of the remaining activities in the LGN. This is the same as the squared length of the difference between the original image and its reconstruction from the integral of V1 activities. The rate of decrease of the error depends on the image being processed. The more an image consists only of sharp edges, the more quickly the representation in terms of edge detectors is formed. Therefore, the simulations proceeded most rapidly for the walsh pattern, then the lena image, and lastly, the random image. Random noise and textures are likely to be represented more efficiently in the brain's image processing by specialized circuits which perhaps model the textures' spectral or statistical properties.

The solution to the Fokker-Planck equation for the image processing network was shown to be gaussian, because the state of the error at $t=0$ is known precisely. The predicted value of the variance was close to that measured for the random-noise image. The variances for other images were very far off, probably because the distribution of the random term of the corresponding Langevin equation is not gaussian in those cases.

It is interesting that the image processing system in the brain might be close to those employed in engineering fields, such as the wavelet and Gabor representations. The pressure for optimization is common to both. The existence of the orientational feature map is unique to the biological processing system. We believe that the feature map arises because the brain is more limited in the complexity of the feature detectors employed than is the engineered system. Therefore more feature detectors of a simpler type must be used, and the feature map is an efficient way of arranging them to cover as much as possible of the space of input images while preserving local continuity. The latter is crucial for the efficiency of the network described here, for the inhibitory feedback will most effectively reduce neural metabolic expenditure when overlapping receptive fields have nearly the same orientation.

\eject
\bibliographystyle{unsrt}
\bibliography{diss}

\begin{figure}[p] 
\begin{center} 
\begin{tabular}[t]{c} 
\subfigure[]{\parbox{0.45\textwidth}{ 
\epsfxsize=0.45\textwidth 
\epsfysize=0.45\textwidth 
\epsffile{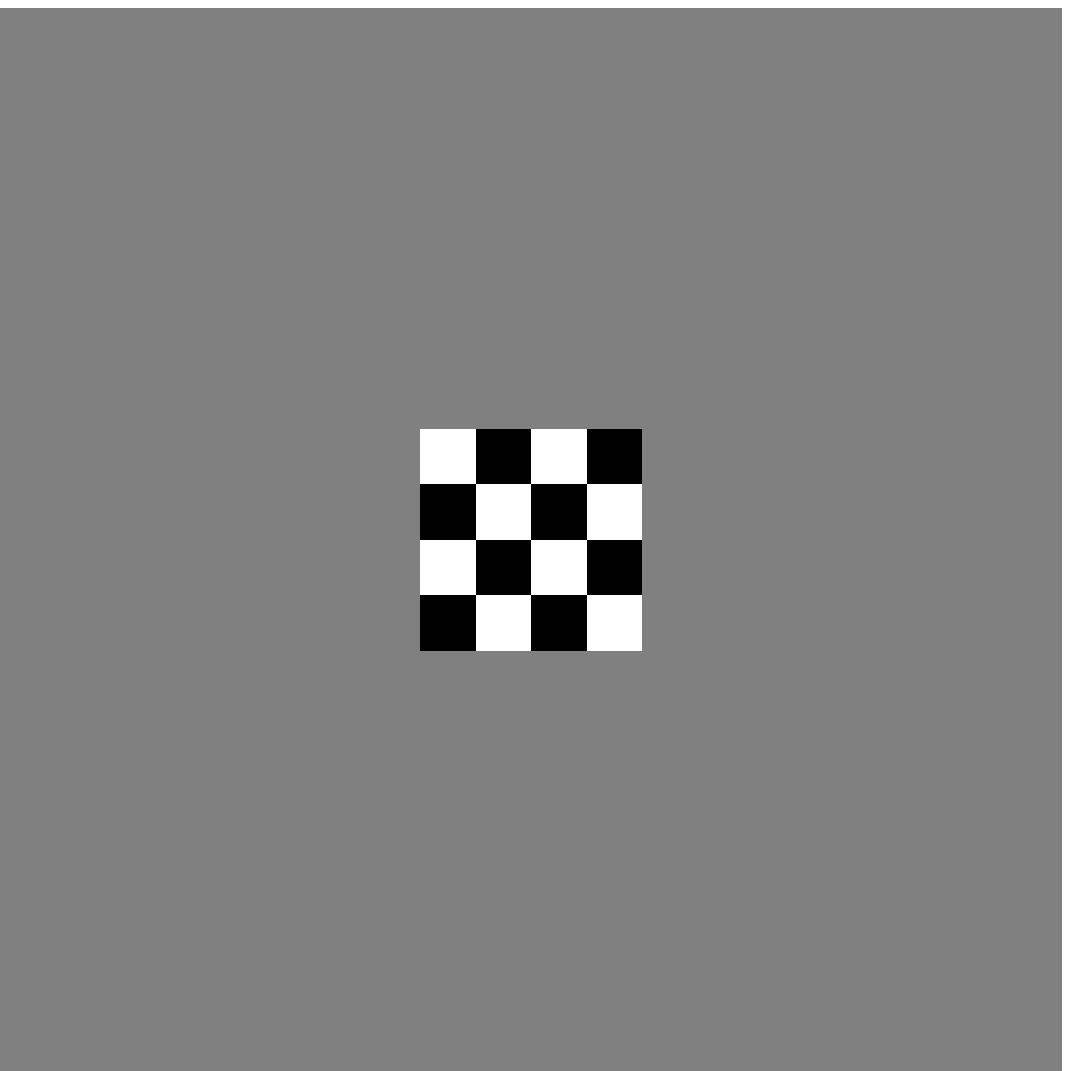}  
} 
} 
\subfigure[]{\parbox{0.45\textwidth}{ 
\epsfxsize=0.45\textwidth 
\epsfysize=0.45\textwidth 
\epsffile{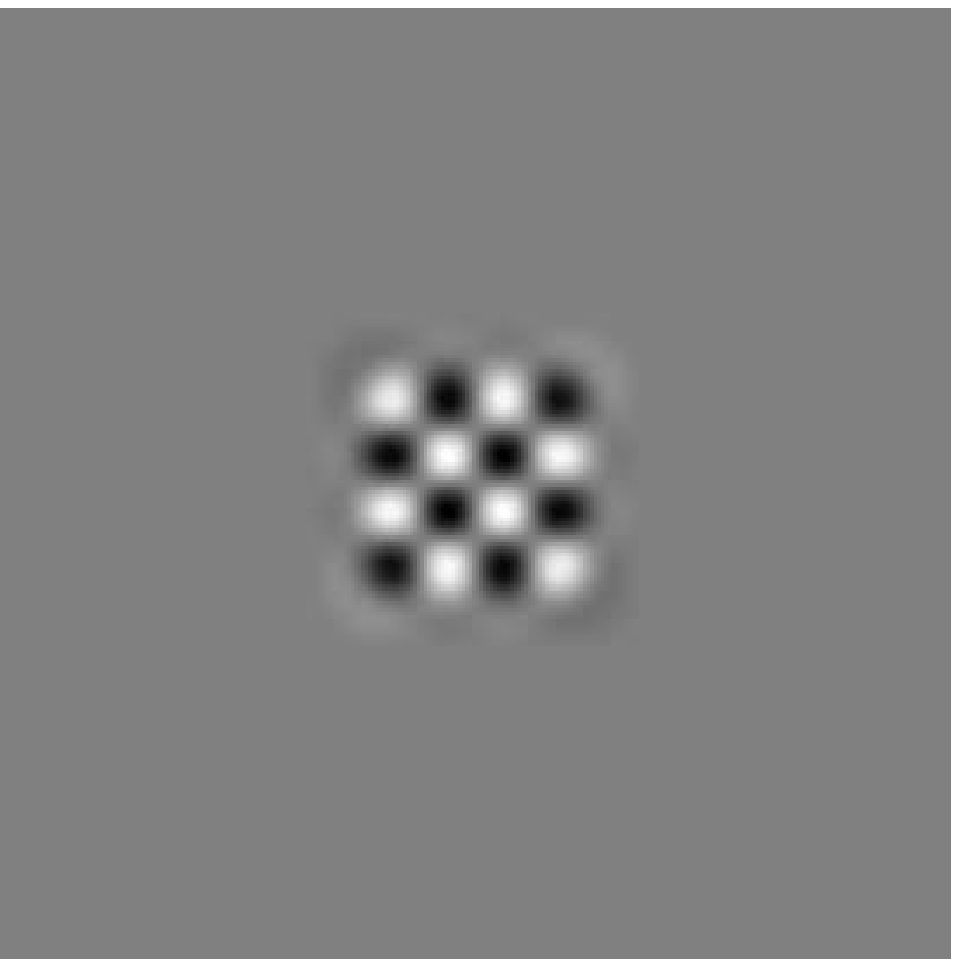}  
} 
} 
\vspace{0.0cm}\\  
\subfigure[]{\parbox{0.45\textwidth}{ 
\epsfxsize=0.45\textwidth 
\epsfysize=0.45\textwidth 
\epsffile{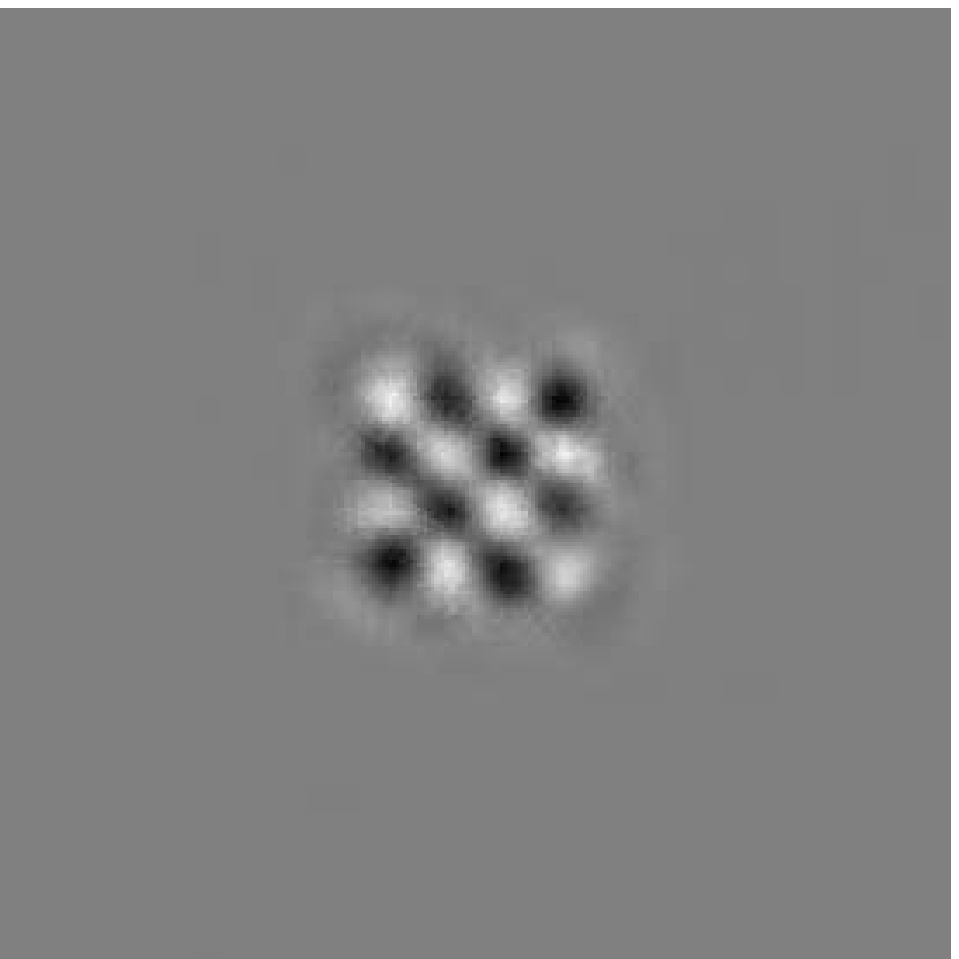} 
} 
} 
\subfigure[]{\parbox{0.45\textwidth}{ 
\epsfxsize=0.45\textwidth 
\epsfysize=0.45\textwidth 
\epsffile{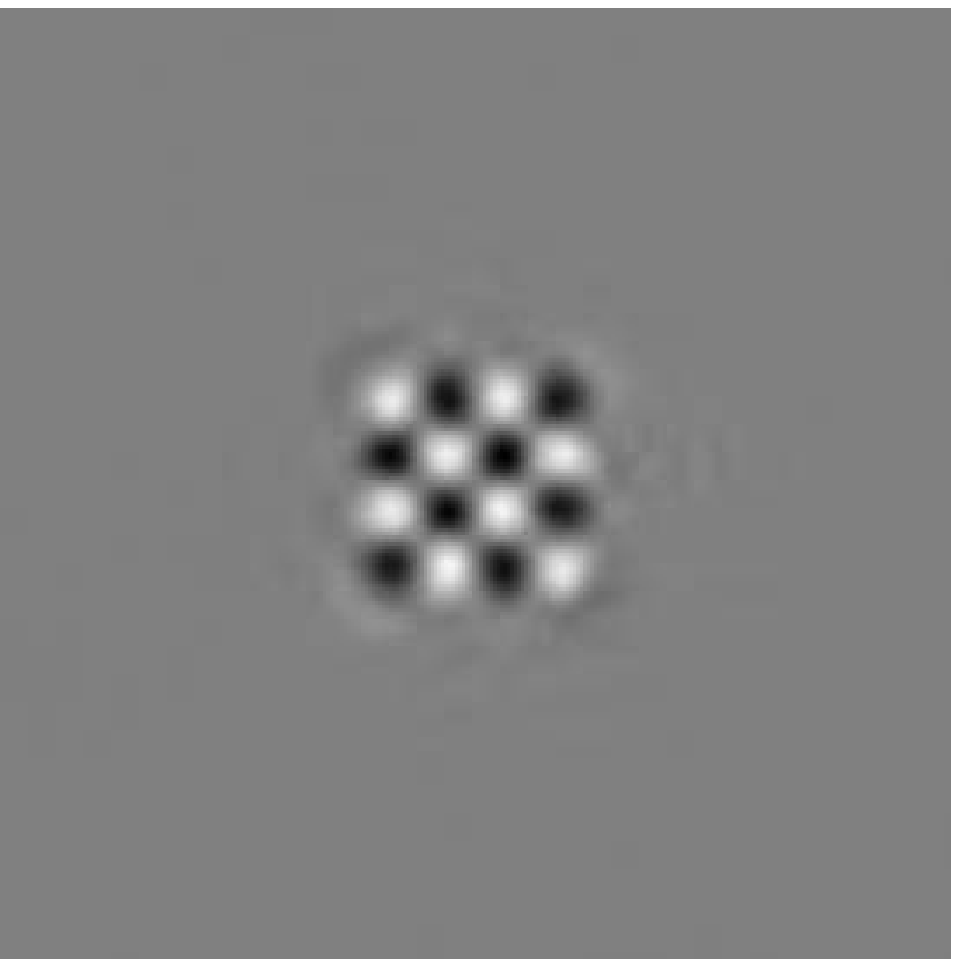}  
} 
} 
\end{tabular} 
\end{center}
\vspace{-0.5cm}
\caption[]{ a) Original walsh pattern (4,4) image. b) Original LGN input for 
walsh image. c) Reconstruction of walsh image after 1 epoch. d) Reconstruction 
of walsh image after 20 epochs.} 
\label{fig:p:walsh44} 
\end{figure}

\eject

\end{document}